\documentclass[useAMS,usenatbib]{mn2e}
\usepackage{amsfonts,amssymb}
\usepackage{graphicx}
\usepackage{ulem}

\def\m{^m\kern-7pt .\kern+3.5pt}
\def\aplt{\ {\raise-.5ex\hbox{$\buildrel<\over\sim$}}\ }
\newcommand{\kms}{\rm km\ s^{-1}}
\newcommand{\ms}{\mbox {$M_{\odot}$}}
\newcommand{\al}{\mbox {$\alpha_{\rm ce} \times \lambda$}}

\begin{document}
\title[]{Population Synthesis for Symbiotic X-ray Binaries}
\author[L\"{u} et al.]{G.-L. L\"{u}$^{1,2}$\thanks{E-mail:
guolianglv@gmail.com (LGL)}, C.-H. Zhu$^{1,2}$, K. A. Postnov$^{3}$,
L. R. Yungelson$^{4}$, A. G. Kuranov$^3$, N.
Wang$^{1}$\thanks{E-mail: na.wang@xao.ac.cn}\\
$^{1}$National Astronomical Observatories / Xinjiang Observatory,
the
Chinese Academy of Sciences, Urumqi, 830011, China\\
$^{2}$School of Physical Science and Technology, Xinjiang
University, Urumqi, 830046,
China \\
$^3$Sternberg Astronomical Institute, Moscow M.V. Lomonosov State University,  13 Universitetski pr., Moscow, 119991, Russia\\
$^{4}$Institute of Astronomy of the Russian Academy of Sciences, 48
Pyatnitskaya Str., Moscow, 119017, Russia\\
}
%\begin{document}

\date{}

\pagerange{\pageref{firstpage}--\pageref{lastpage}} \pubyear{}

\maketitle

\label{firstpage}

\begin{abstract}
Symbiotic X-ray binaries (SyXBs) comprise a rare class of low-mass
X-ray binaries. We study the Galactic SyXBs, which we consider as
detached binaries composed of low-mass giants and wind-fed neutron
star companions, by simulation of the interaction of a magnetized
neutron star (NS) with its environment and utilizing a population
synthesis code. We focus mainly on the parameters that influence
observational appearance of the SyXB: the donor wind velocity (vw)
and the angular momentum distribution in the shell of matter
settling onto NS. We estimate the birthrate of SyXB as $\sim
4.1\times 10^{-5}$ yr$^{-1}$ to $ \sim 6.6\times 10^{-6}$ yr$^{-1}$
and their number in the Galaxy as $\sim$(100 -- 1000). Assumed
stellar wind velocity from cool giants is the input parameter that
influences the model SyXBs population most.

Among known SyXBs or candidate systems, 4U 1954+31 and IGR
J16358-4724 in which NS have very long spin periods may host
quasi-spherically accreting NSs. GX 1+4 has a peculiar long-term
spin behaviour and it may also be a quasi-spherical wind-accreting
source. We cannot identify whether there are wind-fed accretion
disks in 4U 1700+24, Sct X-1, IRXS J180431.1-273932 and 2XMM
J174016.0-290337.
\end{abstract}

\begin{keywords}binaries: symbiotic---pulsar: general---stars:
neutron---X-ray: stars
\end{keywords}
\section{Introduction}
Usually, symbiotic stars  are binaries in which a hot white dwarf
(WD) component burns hydrogen accreted from a cool giant companion
which loses mass via stellar wind  or Roche lobe overflow
\citep{ty76}. Theoretical studies of the population of symbiotic
stars have been published, e.g., by
\cite{Han1995b,Yungelson1995,1996ApJS..105..145I,Lu2006,Lu2008,Lu2009,Lu2011}.

\begin{table*}
%\begin{minipage}{0mm}
  \caption{Parameters of observed symbiotic X-ray binaries. Columns 1 to
       7 list the name of the star, spin period $P_{\rm s}$ and its derivative
       $\dot{P}/P_{\rm s}$, orbital period $P_{\rm orb}$, X-ray luminosity, the distance from the
       Sun, spectral type of companion to NS.  References:
             C97-\citet{Chakrabarty1997}; B97-\citet{Bildsten1997}; H06-\citet{Hinkle2006};
G12-\citet{GonzalezGalan2011};
              C08-\citet{Corbet2008}; M06-\citet{Masetti2006};
              M02-\citet{Masetti2002};
             K07-\citet{Kaplan2007}; M07-\citet{Masetti2007}; N07-\citet{Nucita2007}; P04-\citet{Patel2004};
             P07-\citet{Patel2007}; B06-\citet{Bodaghee2006};
             N10-\citet{Nespoli2010}; L05-\citet{Lutovinov2005}; T06-\citet{Thompson2006};
             F10-\citet{Farrell2010}; M11-\citet{Masetti2011}.
             %M12 -\citet{2012A&A...538A.123M}.
 }
  \tabcolsep0.70mm
  \begin{tabular}{|lcccccccc|}
  \hline\hline
  SyXB&$P_{\rm s}$ (s)&$\dot{P_{\rm s}}/P_{\rm s}$&$P_{\rm orb}$ (days)&$L_{\rm
  X}$ (erg s$^{-1}$)&Distance(Kpc)&Spectral type\\
  \hline
 GX 1+4 &120$^{\rm (C97)}$&transition$^{\rm (B97)}$&1161$^{\rm (H06)}$
 &$10^{35}$---$10^{36}$$^{\rm (G12)}$&4.3$^{\rm (H06)}$&M5 III$^{\rm (C97)}$\\
 4U 1954+31&$\sim 18300$$^{\rm (C08)}$&$-1.4\times10^{-9}$$^{\rm (C08)}$&?
 &$4\times10^{32}-10^{35}$$^{\rm (M06)}$&1.7$^{\rm (M06)}$&M4 III$^{\rm (M06)}$\\
4U 1700+24&?&?&404$^{\rm (M02)}$&$2\times10^{32}-10^{34}$$^{\rm
(K07)}$&0.42$^{\rm (M06)}$
&M2 III$^{\rm (M02)}$\\
Sct X-1&113$^{\rm (K07)}$&$3.9\times10^{-9}$$^{\rm
(K07)}$&?&$2\times10^{34}$$^{\rm (K07)}$
&$\geq4^{\rm (K07)}$&Late K/early M I-III$^{\rm (K07)}$\\
IGR J16194-2810 &?&?&?&$\leq 7\times10^{34}$$^{\rm
(M07)}$&$\leq3.7^{\rm (M07)}$
&M2 III$^{\rm (M07)}$\\
IRXS J180431.1-273932&494$^{\rm
(N07)}$&?&?&$\leq6\times10^{34}$$^{\rm (N07)}$
&10$^{\rm (N07)}$? &M6 III$^{\rm (N07)}$\\
IGR J16358-4724&5850$^{\rm (P04)}$&$3.1\times10^{-8}$$^{\rm
(P07)}$&?&$3\times10^{32}-3\times10^{36}$$^{\rm (P07)}$ &5-6;
12-13$^{\rm (L05)}$ &K-M III$^{\rm
(N10)}$\\
IGR J16393-4643&912$^{\rm (B06,T06)}$&$1.0\times10^{-11}$$^{\rm
(N10)}$&50.2$^{\rm (N10)}$&?
&$\sim 10$$^{\rm (B06)}$ &K-M III$^{\rm (N10)}$ \\
2XMM J174016.0-290337&626$^{\rm (F10)}$&?&?&$\sim3\times10^{34}$$^{\rm (F10)}$&$\sim 8.5$$^{\rm (F10)}$ &K1 III$^{\rm (F10)}$ \\
CGCS 5926&?&?&$\sim$3000$^{\rm (M11)}$&?&$5$$^{\rm (M11)}$ &C$^{\rm (M11)}$ \\
\hline\hline
 \label{tab:syxb}
\end{tabular}
\end{table*}

Recently, a small but rapidly growing subclass of symbiotic stars
gained attention, in which accreting component is a neutron star
(NS). All these systems are hard X-ray emitters \citep{Murset1997}.
They  are low-mass X-ray binaries, with possible exception of Sct
X-1 \citep{Kaplan2007}, for which a high-mass solution also is
viable.
\cite{Masetti2006}  dubbed these systems symbiotic X-ray binaries
(SyXBs). Currently, 10 binaries are classified as  SyXBs or
candidate systems (Table \ref{tab:syxb}).

Remarkable features of SyXBs are their  orbital periods, among the longest determined for
low-mass X-ray binaries \citep{Liu2007} and long
NS spin periods.
With a 18300\,s spin period,  4U~1954+31 is the slowest accretion-powered
NS known to date \citep{Masetti2006}. Long spin periods of NS in SyXBs are not
easy to understand assuming the standard disk accretion. In that
case, equilibrium NS spin periods would require magnetar-like
magnetic fields of NSs. However, long periods find natural
explanations if accretion onto NS proceeds quasi-spherically. The
quasi-spherical accretion onto NS was reconsidered in the recent
paper by \cite{Shakura2011}. It was shown that at small X-ray
luminosities, $L_x<4\times 10^{36}$~erg/s, the subsonic settling
accretion regime from the stellar wind onto NS takes place, during which an effective removal of
angular momentum from the NS magnetosphere can take place through hot
convective shell formed above the magnetosphere. In this accretion
regime, equilibrium spin periods of NS can be very long even for
the standard values of the NS magnetic fields $\sim
10^{12}-10^{13}$~G. This theory was applied for explanation of
the observed features of low-luminosity X-ray pulsars, including
long-term spin-down of NS and spin-luminosity correlations in GX 1+4 \citep{GonzalezGalan2011}, the
5-h spin period of 4U~1954+31 \citep{Marcu2011}, the NS spin and luminosity
behaviour in X Per \citep{Lutovinov_ea2012}, etc.

Clearly, the model of the Galactic population of SyXBs needs a  detailed treatment
of accretion process onto NSs, to which we pay particular attention in the present paper.
We focus on the population synthesis for SyXBs. We simulate zero-age
population of NSs with companions and then follow evolution of the
population with special accent on the  spin  and magnetic field of
NSs interacting with the matter lost by their companions. In \S2
we present our assumptions and describe some details of the modelling
algorithm. In \S3 we present a detailed  example of the evolution of
a NS interacting with its companion. In \S4 the properties of the
model population of SyXBs are presented  and individual observed SyXBs
are discussed. Conclusions follow in \S5.

\section{The Models}

For the simulation of binary evolution, we use rapid binary star
evolution code BSE \citep{Hurley2002} with updates by
\cite{Kiel2006}.
 If any input parameter is not specially mentioned,
%it is
its default value is taken from these papers.
%taken as default value in \cite{Hurley2002} and \cite{Kiel2006}.
The code is extended with an algorithm for the treatment of spin
evolution of a magnetized NS, as described in \S\S \ref{sec:class},
\ref{sec:spin}, \ref{sec:mfe}.

%\subsection{Initial input parameters}

Like in the main case considered in our study of symbiotic stars
with white dwarf components \citep{Lu2006},
 we use \citet{1979ApJS...41..513M} initial mass-function for primary components,
a flat distribution of initial mass ratios of components
\citep{1989Ap.....30..323K,1994A&A...282..801G}. We assume that all
binaries have initially circular orbits. After a supernova
explosion, new parameters of the orbit are derived using standard
formulae, \citep[e. g., ][]{Hurley2002}.

The model is normalized to formation of one binary with $M_1 \geq
0.8 M_\odot$\ per year \citep{Yungelson1993}. We use $10^7$ binary
systems in the Monte-Carlo simulations. This gives a statistical
error $\leq$5\% for the number of Galactic SyXBs.

Below, we only mention several
specific assumptions used in the code.

\subsection{Kick Velocity}
\label{sec:kick}

Nascent NS receive an additional velocity  (``kick'') due to a
still enigmatic process that violates spherical symmetry during the collapse of a massive star
or later.
The kicks have dichotomous nature,
as it was suggested quite early
by \citet{Katz1975} and later
confirmed by, e. g.,
\citet{Hartman1997,Pfahl2002}. Observationally, the kick is not well
constrained due to numerous selection effects. Currently, high
($\sim100\,\rm km\ s^{-1}$) kicks are associated with NS originating
from core-collapse supernovae, while low kicks ($\sim 10\rm km\
s^{-1}$) --- with NS born in electron-capture supernovae and
accretion-induced collapses (see, e. g., \citet{Ivanova2008} for
references and discussion).

Following \citet{Podsiadlowski2004}, we assume that core collapses
are experienced by stars with ZAMS
mass $ M/M_\odot \geq 11$.
For NSs born via core-collapse we apply
Maxwellian distribution of kick velocity $v_{\rm k}$
\begin{equation}
P(v_{\rm k})=\sqrt{\frac{2}{\pi}}\frac{v^2_{\rm k}}{\sigma^3_{\rm
k}}e^{-v^2_{\rm k}/2\sigma^2_{\rm k}}.
\end{equation}
We use velocity dispersion $\sigma_{\rm k}= 190$ km s$^{-1}$, which
is consistent with the data on pulsar proper motions
\citep{Hansen1997}. Variation of $\sigma_{\rm k}$ between 50 and 200
km s$^{-1}$, introduces an uncertainty $\lesssim 3$ in the birthrate
of low- and intermediate-mass X-ray binaries \citep{Pfahl2003}.
Since in this paper we focus on the physical parameters of close
binaries that mostly affect  observational appearance of NS ---
their equilibrium spin period and X-ray luminosity,  we do not
discuss the effects of $\sigma_{\rm k}$ on SyXB's population.

%\subsubsection{Electron capture supernovae and \\
%accretion-induced collapse}

Electron-capture supernovae (ECS) and accretion-induced collapses
(AIC) are associated with formation of ONeMg cores of stars or white
dwarfs.
%in which
Collapse is triggered by electron captures on
$^{20}$Ne and $^{24}$Mg \citep{Miyaji1980}. ECS occur in
single stars, while AIC in white dwarfs in close binaries. The range
of stars that form ONeMg cores or dwarfs depends on subtle details
of the treatment of rotation, mass loss, mixing etc. and also depends on
stellar evolution code applied \citep[see, e. g., discussion
in][]{Podsiadlowski2004}. Following \cite{Kiel2008}, we assume that
ECS occurs if  stellar helium core mass is  $1.4 \leq M_{\rm He}/M_\odot \leq
2.5$. Progenitors of these stars have ZAMS mass between $\sim$ 8.0
and 11.0 $M_\odot$ \citep{Hurley2000}. Accretion-induced collapses
happen when accreting ONeMg WD reach the Chandrasekhar limit. Progenitors of ONeMg
WDs have ZAMS mass between $\sim$ 6.3 and 8.0
$M_\odot$ \citep{Hurley2000}.

Considering the problem of retention of NSs in globular clusters,
\cite{Pfahl2002}  suggested that NSs born in ECS and AIC have
velocity dispersion lower than $50\,\kms$. In this study we assume it equal to  20 $\kms$.

Response of  ONeMg dwarfs to accretion  is treated in the same way as for
CO WD \citep[see for the details][]{Lu2009}.

\subsection{Mass Transfer and Angular Momentum Loss}
\label{sec:mstr}

Mass transfer may occur by accretion of the matter lost via stellar wind
or due to Roche-lobe overflow. If the system is detached, we apply
\citet{Bondi1944} accretion formula in which  mass-accretion rate
greatly depends on the wind velocity $v_{\rm w}$. Determination of
$v_{\rm w}$\ is difficult. According to \citet{Harper1996}, the main
characteristic of the cool winds of evolved K and early M giants is
that their terminal velocities are lower than the surface escape
velocity, typically, $v_{\rm w}\leq1/2v_{\rm esc}$.
As the wind velocity is very important parameter for accretion, we
set $v_{\rm w}=1/2v_{\rm esc}$ and $v_{\rm w}=2v_{\rm esc}$ in
different simulations.

For the case of RLOF,
the critical issue is whether mass loss occurs on dynamical time
scale and common envelope (CE) forms. This depends on the mass ratio
of components $q=M_{\rm d}/M_{\rm a}$. Following \citet{Hurley2002},
we assume that  condition for formation of CE (if the accretor is a
main-sequence star) is $q>4$ for the donors in the Hertzsprung gap.
If the donor is a giant, critical condition is given by $q>q_{\rm
cr}$, where
\begin{equation}
q_{\rm cr}=0.362+\frac{1}{3\times(1-M_{\rm c}/M)}, \label{eq:qcrit}
\end{equation}
where $M_{\rm c}$ is the mass of the stellar degenerate core
\citep{Webbink1988}.

For common envelopes, we apply \citet{Webbink1984} equation for
energy balance, as modified by \citet{dek90} by adding a numerical
``structure''  factor $\lambda$,  meant to describe the dependence
of binding energy of the donor on the density distribution.  Then,
final separation of the components after CE-stage depends on the
product of $\lambda$ and ``common envelope efficiency'' $\alpha_{\rm
ce}$ (the fraction of binary binding energy which is spent to expell
common envelope). We assume \al=0.5. \citet{Pfahl2003} have shown
that such high value of \al\ is necessary for explanation of the
birthrate of Galactic population of LMXB. Reduction of \al\ may
result in decrease of the birthrate of SyXB due to increase of the
number of mergers.

%{\bf During a CE phase, binary undergoes a dynamical spiral-in. It
%is generally assumed that the energy of binary system is
%conventional and the orbital energy of the binary is used to expel
%the envelope of the donor with an efficiency $\alpha_{\rm ce}$. In
%the theoretical calculation, the dynamical spiral-in are affected by
%the ¡®combined¡¯ parameter $\alpha_{\rm ce}$$\lambda_{\rm ce}$, in
%which $\lambda$ parameterizes the structure of donor's envelope.
%\cite{Pfahl2003} discussed the effects of the  parameter
%$\alpha_{\rm ce}$$\lambda_{\rm ce}$ on LMXB's populations.
%Increasing $\alpha_{\rm ce}$$\lambda_{\rm ce}$ from 0.1 to 0.5, they
%found that the total number of the Galactic LMXBs increases to $\sim
%10^5$ from $\sim 10^3$. The parameter $\alpha_{\rm ce}$$\lambda_{\rm
%ce}$ has a great effect on LMXB's populations. However, similar to
%\S \ref{sec:kick}, we do not investigate the effect of the parameter
%$\alpha_{\rm ce}$$\lambda_{\rm ce}$ on SyXB's populations. In this
%work, we take $\alpha_{\rm ce}$$\lambda_{\rm ce}=0.5$. }

In the treatment of angular momentum loss we follow original BSE code with only exception:
for magnetic braking we replace  Eq.~(50) of \cite{Hurley2002}
by a formula from  \cite{Verbunt1981} with $\lambda=1$ and $k^2=0.1$\footnote{The reason for this change is  the following.
\cite{Hurley2002} use a modification of magnetic braking law parametrization suggested by \citet{1983ApJ...275..713R}.
Then, BSE produces a long-living population of systems with giant donors with mass $\simeq 0.1$\,\ms,
contradicting observations.}.

\subsection{Evolution of Neutron Stars}
\label{sec:class}

The regime of interaction of a rotating magnetized
NS (single or in a binary system) with surrounding matter has been
recognized to be the most important astrophysical aspect of its
evolution, see, e. g., \citet{Pringle1972,Illarionov1975,Ghosh1978,Ghosh1979a,Ghosh1979b,Lovelace1995,Lovelace1999}).
From the point of view of population synthesis, a convenient way
of describing NS evolution was elaborated by \cite{Lipunov1992}.
Progress in the 3D MHD simulations of the interaction of a rotating
magnetized NS with accreting plasma
% (e.g. Romanova et al [give
%refs!])
generally confirmed the basic ideas
\citep{Romanova2002,Romanova2003,Romanova2004,Romanova2005}.

The regime of interaction of a rotating magnetized NS with its
environment can be determined by relations between four characteristic radii:
\begin{itemize}
\item the radius of the light cylinder $R_{\rm l}={\rm c}/\omega$ where c is
      the speed of light and $\omega$ is the spin frequency of NS;
\item the radius of the gravitational capture (the Bondi radius)
$$
R_{\rm G}=\frac{2GM_{\rm NS}}{v^2_{\rm \infty}},
$$
     where $M_{\rm NS}$ is the mass of NS and $v_{\rm \infty}$ is
     the velocity of surrounding matter;
\item the corotation radius $R_{\rm c}=(GM/\omega^2)^{1/3}$;
\item the radius $R_{\rm st}$ where the flow of accreted
     matter (free-fall or accretion disk)  is stopped due to  interaction with the NS magnetosphere.

\end{itemize}

Depending on the relations between the above four radii,  NS can
be
 in several basic evolutionary states \citep{Lipunov1992}:
\begin{itemize}
  \item Ejector state [$R_{\rm st}>\max(R_{\rm G}, R_{\rm l})$],
  in which the pressure of relativistic particle  wind
  is sufficient to keep the stellar wind plasma of the
  companion away from the NS magnetosphere. For example, ordinary radio
pulsars form a subclass of ejectors (additional physical conditions may be
required for effective relativistic plasma creation);
  \item Propeller state [$R_{\rm c}<R_{\rm st}<\max(R_{\rm G}, R_{\rm
  l})$],
  in which
  the relativistic wind pressure cannot
  prevent the infalling matter from interaction with the
  magnetosphere, while the fast rotating magnetic field prevents the matter
  from stationary accretion onto the NS surface;
  \item Accretor state ($R_{\rm st}<R_{\rm c}$), in which accretion onto magnetized NS is
   centrifugally allowed,
   and the NS magnetic field channels the accreting material toward the
   magnetic poles of the NS.
\end{itemize}

We shall assume that $R_{\rm st}$ is equal to the Alfv$\acute{\rm e}$n
radius $R_{\rm A}$. For disk and quasi-spherical supersonic
accretion (see below)
\begin{equation}
R_{\rm A}=(\mu^2/(\dot{M}_{\rm NS}\sqrt{2GM_{\rm NS}}))^{2/7},
\label{eq:ra}
\end{equation}
where $\dot{M}_{\rm NS}$ is the accretion rate onto NS,  magnetic
dipole moment $\mu=B_{\rm NS}R^3_{\rm NS}/2$, and $B_{\rm NS}$ and
$R_{\rm NS}$ are the magnetic field and the radius of NS,
respectively.

However, Eq. (\ref{eq:ra}) is valid only if the accretion rate is
not too high and the accretion luminosity is below the Eddington
limit.
For the supercritical accretion via a disk, \citet{Shakura1973}
suggested that the accretion rate in the disk begins to fall
monotonically from certain radius $R=R_{\rm s}$ (the spherization
radius). The super-Eddington accretion onto magnetized NSs was
studied in detail by \cite{Lipunov1982} and summarized in
\cite{Lipunov1987}. The spherization radius $R_{\rm s}$ is
approximately given by
$$
R_{\rm
s}=\frac{\kappa}{4\pi \rm c}\dot{M}_{\rm NS},
$$
where  $\kappa$ is
the opacity of the accreting matter. When $\dot{M}_{\rm NS}$ is higher
than the Eddington accretion rate,  NSs can accrete
via the disk at a rate
$$\dot{M}_{\rm
NS}^{\rm S}=\frac{R_{\rm A}}{R_{\rm s}}\dot{M}_{\rm NS}.
$$
The excess of the matter is expelled in the form of a wind from
a supercritical accretion disk and carries away the specific orbital
angular momentum of the NS. In the supercritical regime  the
Alfv\'{e}n radius is determined as (see \cite{Lipunov1987})
\begin{equation}
R_{\rm SA}=\left(\frac{\mu^2\kappa}{4\pi {\rm c} \sqrt{2GM}}\right)^{2/9}.
\end{equation}
Similarly, in the case of  super-Eddington accretion, depending on
the relations between
 $R_{\rm l}$, $R_{\rm G}$, $R_{\rm c}$, and $R_{\rm A}$,  NS can
be in super-ejector (SE), super-propeller (SP)  and super-accretor
(SA) states \citep{Lipunov1987}.

For a NS, "evolution" means that its basic parameters which
determine the interaction with the surrounding medium change. First
of all, this concerns NS spin and magnetic field. The effect of the
NS mass increase in the accretion state is generally less important,
unless a hypercritical accretion (especially inside common
envelopes) is allowed. Then the accretion-induced formation of a
black hole may occur (see, e. g., \cite{Brown2004} for more details).

\subsubsection{Spin evolution}
\label{sec:spin}

Spin evolution (spin-up or spin-down) of a NS in a binary system can
be conveniently described by an angular momentum conservation
equation
\begin{equation}
\frac{{\rm d}I\omega}{{\rm d}t}=K_{\rm su}-K_{\rm sd}\,,
\label{eq:torq}
\end{equation}
where $I$ is  NS momentum of inertia, $K_{\rm su}$ and $K_{\rm sd}$
are spin-up and spin-down torques, respectively. Their forms are
different in different evolutionary states and also depend on the
mode of accretion, i.e.,   whether accretion disk is present or
accretion proceeds quasi-spherically.  At small accretion rates, spin
evolution of a NS in a binary system proceeds almost
 identically to that of a single NS, so we neglect interaction with matter for
 $\dot M <10^{-15} M_\odot\ {\rm yr}^{-1}$ and treat  NS
spin evolution as in the ejector state.

In the ejector state, $K_{\rm su}=0$ and  NS spins down due to
%\sout{magneto-dipole \citep{Ostriker1969} or}
current braking \citep{Beskin1993}. To within a numerical factor
depending on the angle $\xi$ between the spin and magnetic dipole
axes, the spin-down torque can be written, using the light-cylinder
radius $R_{\rm l}$, as $K_{\rm sd}=2\mu^2/(3R^3_{\rm l})$. The
secular change of the angle $\xi$ is beyond the scope of our
consideration.

In the propeller state, accretion is centrifugally prohibited,
$K_{\rm su}=0$, and the spin-down torque can be written using
 the
magnetospheric radius in the form $K_{\rm sd}=k_{\rm
t}\mu^2/R^3_{\rm A}$. The numerical factor, $k_{\rm t}$ is
model-dependent but is $\sim 1$\citep{Shakura2011}.

In the accretor state, both  $K_{\rm su}$  and $K_{\rm sd}$
are determined by the  geometry of the accretion flow, as specified below.

{\it Roche lobe overflow.} If accretion onto the compact
star occurs via  Roche lobe overflow through the vicinity of the
inner Lagrangian point, an accretion disk is formed \footnote{Unless the
mass-donating component turns out to be inside the magnetosphere of
the compact object; this case can be relevant for the accretion onto
a magnetized white dwarf, as in AM Her systems}. The spin-up torque
in both accretor and super-accretor states is determined by the
specific angular momentum at the inner disk radius and is $K_{\rm
su}= \dot{M}_{\rm NS}\sqrt{GM_{\rm NS}R_{\rm A}}$. The spin-down
torque for disk accretion in the first approximation
can be written in all cases using the
 corotation radius  $R_{\rm c}$ as
$K_{\rm sd}=(1/3)\mu^2/R^3_{\rm c}$.

The total torque exerted on a NS in the accretor or super-accretor state
is then
\begin{equation}
K=\dot{M}_{\rm NS}\sqrt{GM_{\rm NS}R_{\rm A}}-1/3 \mu^2 R^{-3}_{\rm c}.
\label{eq:dtor}
\end{equation}
The competitive action of spin-up and spin-down torques, on average,
diminishes the total torque acting on NS, and an equilibrium state
of NS is reached.  The equilibrium NS spin period is then
\citep{Lipunov1988}
\begin{equation}
P_{\rm s}^{\rm eq}=5.72M_{\rm
NS}^{-5/7}\dot{M}_{16}^{-3/7}\mu_{30}^{6/7}\ {\rm s,} \label{eq:deq}
\end{equation}
where $\mu_{30}=\mu/(10^{30}{\rm G\ cm^3})$ is the NS dipole
magnetic momentum, and $\dot{M}_{\rm 16}=\dot{M}_{\rm
NS}/(10^{16}{\rm g/s})$ is the accretion rate.

For super-Eddington accretion, the NS equilibrium period is found to
be
\begin{equation}
P_{\rm s}^{\rm eq}=1.76\times10^{-1}\mu_{30}^{2/3}M_{\rm NS}^{-2/3}\
{\rm s}.
 \label{eq:seq}
\end{equation}

{\it Wind-fed accretion.} In the wind-fed X-ray binaries, the
physical condition for formation of an accretion disk is $j_{\rm
a}>j_{\rm K}(R_{\rm A})$, where $j_{\rm a}=k_{\rm w}\Omega_{\rm b}
R^2_{\rm G}$ is  specific angular momentum of the captured
stellar wind matter, $\Omega_{\rm b}=2\pi/P_{\rm orb}$ is
orbital frequency, $k_{\rm w}$ is a numerical coefficient, which  we set
 to 0.25 after \cite{Illarionov1975}, $j_{\rm K}(R_{\rm
A})=\sqrt{GM_{\rm NS}R_{\rm A}}$ is the Keplerian angular momentum
at the NS magnetosphere. If an accretion disk forms, the case of
wind-fed disk accretion is realized, otherwise, a regime of
quasi-spherical accretion  is established. We shall
assume that the wind-fed disk accretion is similar to the disk
accretion via Roche lobe.

As discussed by \cite{Shakura2011},
there can be
two different regimes of quasi-spherical accretion.
First, if the gas
heated up in the bow shock cools down rapidly, the matter falls toward NS
supersonically. This happens if accretion rate is fairly high (above
$\sim 4 \times 10^{16}$~g/s). In this regime, all gravitationally
captured matter eventually reaches the NS surface. Spin
evolution of NS is determined by the sign of the specific angular
momentum of captured matter, which can be positive (prograde) or
negative (retrograde), and NS can spin-up or spin-down,
respectively. Numerical simulations \citep{Ho_ea1989} show that the
sign of the specific angular momentum of the captured matter can
alternate; here, however, we shall consider only the prograde case,
i.e.  only  NS spin-up during the supersonic accretion with
$K_{su}=\dot M_{NS} \sqrt{GMR_A}$. To within a factor of 2,  the
magnetospheric radius $R_A$ in this regime coincides with that for
disk accretion [Eq.(\ref{eq:ra})], and here we shall ignore the
difference (but see below).

Second, if accretion rate onto NS is smaller than $4\times
10^{16}$~g/s, shocked gas remains hot, a quasi-static extended shell
forms around the NS magnetosphere, and accretion proceeds in the
subsonic settling regime. In this regime, the actual accretion rate
onto NS $\dot M_{\rm NS}$ can be lower than the capture rate by NS,
and its value is essentially determined by the ability of plasma to
enter the NS magnetosphere via instabilities; however, in this paper
we shall assume that in this regime all matter captured by the NS
from the stellar wind ultimately accretes onto the NS, as in the
case of supersonic accretion. This assumption implies that the X-ray
luminosity at the model SyXB stage can be higher than in reality,
leading to shorter equilibrium NS spin periods. As shown by
\cite{Shakura2011}, large-scale convective motions set in the shell,
and the latter mediates angular momentum transfer to or from the NS
magnetosphere depending on X-ray luminosity. The NS
spin-up/spin-down equation can be written in the form given by
Eq.(\ref{eq:torq}), with the spin-down torque
\begin{equation}
K_{\rm sd}=Z_2 \dot{M}_{\rm NS} \omega R_{\rm A}^2 \label{eq:ksd}
\end{equation}
and the spin-up torque
\begin{equation} K_{\rm
su}=Z_1 \dot{M}_{\rm NS} \Omega_{\rm b} R_{\rm
G}^{2}\left(\frac{R_{\rm A}}{R_{\rm G}}\right)^{2-n}\,.
\label{eq:ksu}
\end{equation}
Here, the dimensionless coefficient $Z_1$ determines the
plasma-magnetosphere coupling and it is a function of the accretion rate
$\dot M_{NS}$
and NS dipole magnetic moment $\mu$, $Z_2=Z_1-2/3$. The value of the index $n$ reflects
different rotational distributions  of matter inside the shell ($\omega(R)\sim
R^{-n}$); it depends on the treatment of (generally, anisotropic)
turbulent viscosity responsible for angular momentum transport in
the shell. The analysis of observed X-ray pulsars
\citep{Shakura2011} suggests that an iso-angular momentum
distribution with $n=2$ is most likely, but quasi-Keplerian law with
$n=3/2$ cannot be excluded as well. Therefore, in our modeling we
shall consider both quasi-Keplerian ($n=3/2$) and iso-angular
momentum distributions ($n=2$) as representative cases.

It is important to note that the value of the  Alfv$\acute{\rm e}$n radius itself
in the regime of settling accretion is different from that in the
case of disk accretion or supersonic quasi-spherical accretion (Eq.
(\ref{eq:ra}) above):
\begin{equation}
R_{\rm A}\approx 10^9 \left(\frac{\mu_{30}^3}{\dot
M_{16}}\right)^{2/11}\left(\frac{M_{\rm
NS}}{1.5M_\odot}\right)^{-2/11}~{\rm cm}\,.
\end{equation}
%(\textbf{Here and below in numerical coefficients we assumed the NS mass to be
%$M_{NS}=1.5 M_\odot$}).
After inserting this expression into the formulae for spin-up/spin-down torques
and using the value of $Z_1$ from \cite{Shakura2011}, the NS spin
evolution equation reads:
\begin{equation}
\label{eq:qsspinev}
I\dot \omega=A\dot M^{\frac{3+2n}{11}} - B\dot M^{3/11}\,,
\end{equation}
The accretion-rate independent coefficients $A$ and $B$ are (in CGS units):
\begin{equation}
\begin{array}{ll}
A\approx &5.33\times 10^{31} (0.034)^{2-n}K_1
\mu_{30}^{\frac{13-6n}{11}}v_8^{-2n}\\&\times\left(\frac{P_{\rm
orb}}{10\hbox{d}}\right)^{-1}
\left(\frac{M_{\rm NS}}{1.5M_\odot}\right)^{47/22-13/11(2-n)},\\
\end{array}
\end{equation}
\begin{equation}
B\approx 5.4\times
10^{32}K_1\mu_{30}^{\frac{13}{11}}\left(\frac{P_{\rm
s}}{100\hbox{s}}\right)^{-1}\left(\frac{M_{\rm
NS}}{1.5M_\odot}\right)^{-5/22}.
\end{equation}
Here $v_8=v_w/(1000$~km/s) is the relative stellar wind velocity,
$K_1$ is a dimensionless numerical factor which  differs for
different systems; we shall assume $K_1=40$, which is the typical
value found from the analysis of observed X-ray pulsars
\citep{Shakura2011}.

In the equilibrium state (when the total torque exerted on the NS vanishes,
$K_{su}+K_{sd}=0$), the equilibrium spin period of a  NS in the
settling accretion state is independent of the source-specific
coefficient $K_1$ and it is given by
\begin{equation}
\begin{array}{ll}
P_{\rm s}^{\rm eq}\approx &1000\cdot (0.034)^{n-2}
\mu_{30}^\frac{6n}{11}v_8^{2n}\dot M_{16}^{-\frac{2n}{11}}\\
&\times\left(\frac{P_{\rm orb}}{10\hbox{d}}\right)\left(\frac{M_{\rm
NS}}{1.5M_\odot}\right)^{13/11(2-n)-26/11}~{\rm s}.\\
%\frac{Z_2}{Z_1}P_{\rm orb}\left(\frac{R_{\rm
%A}}{R_{\rm G}}\right)^{2} \ {\rm or} \  P_{\rm s}^{\rm
%eq}=\frac{Z_2}{Z_1}P_{\rm orb}\left(\frac{R_{\rm A}}{R_{\rm
%G}}\right)^{3/2}
\end{array}
\label{eq:weq}
\end{equation}
In our calculations, we take $M_{NS}=1.5 M_\odot$ in Eqs. (11),
(13)-(15). Among other factors, this one appears to be known quite
well. In all formulae we omit some lengthy factors describing
anisotropic turbulence in the quasi-spherical envelope; taking them
into account would change the result to within less than 10 per
cent.

\begin{table*}
%\begin{minipage}{0mm}
\caption{Spin-up $K_{\rm su}$ and spin-down $K_{\rm sd}$ torques
acting on the rotating magnetized NS at basic evolutionary states,
and the corresponding equilibrium periods derived from the equation
$K_{\rm su}-K_{\rm sd}=0$. We use units
$\mu_{30}=\frac{\mu}{10^{30}{\rm G\ cm^3}}$,
$\dot{M}_{16}=\frac{\dot{M}_{\rm NS}}{10^{16}{\rm g/s}}$, and
$R_{\rm A, 8}=\frac{R_{\rm A}}{10^{8}{\rm cm}}$.
               }
  \tabcolsep0.70mm
  \begin{tabular}{|l|c|c|c|}
  \hline\hline
State & $K_{\rm su}$& $K_{\rm sd}$& $P_{\rm eq}$\\
  \hline
E       &  ---  & $\frac{2}{3}\frac{\mu^2}{R_{\rm l}^3}$   & ---\\
\hline
P       & ---   &  $k_{\rm t}\frac{\mu^2}{R_{\rm A}^3}$ &---\\
\hline
A, disk  & $\dot M_{\rm NS}\sqrt{{\rm G}MR_{\rm A}}$   & $ \frac{1}{3}\frac{\mu^2}{R_{\rm c}^3}$ & subcritical:\\
&&&$ 5.72M_{\rm NS}^{-5/7}\dot{M}_{16}^{-3/7}\mu_{30}^{6/7}\ {\rm s}
$ \\
&&& supercritical:\\
&&& $ 1.76\times10^{-1}\mu_{30}^{2/3}M_{\rm NS}^{-2/3}\ {\rm
s}$ \\
A, quasi-sph., & $\dot M_{\rm NS}\sqrt{GMR_{\rm A}}$ & - & - \\
supersonic ($\dot M_{\rm NS}\ge 4\cdot 10^{16}$~g/s)\\

\hline A, quasi-sph., & $Z_1 \dot{M}_{\rm NS} \Omega_{\rm b} R_{\rm
G}^{2}\left(\frac{R_{\rm A}}{R_{\rm G}}\right)^{2-n}$\ \  \ \ \   &
$Z_2\dot {M}_{\rm NS} \omega R_{\rm A}^2$ \ \ \ \ \ \ &
$ 1000\cdot (0.034)^{n-2}\hbox{[s]}\, \mu_{30}^\frac{6n}{11}v_8^{2n}\dot M_{16}^{-\frac{2n}{11}}$ \\
%\frac{Z_2}{Z_1}P_{\rm orb}\left(\frac{R_{\rm A}}{R_{\rm
%G}}\right)^{2} {\rm or} \frac{Z_2}{Z_1}P_{\rm orb}\left(\frac{R_{\rm
%A}}{R_{\rm G}}\right)^{3/2}$\\
settling ($\dot M_{\rm NS}<4\cdot 10^{16}$~g/s)&&&
$\times\left(\frac{P_{\rm orb}}{10\hbox{d}}\right)\left(\frac{M_{\rm
NS}}{1.5M_\odot}\right)^{13/11(2-n)-26/11}~{\rm s}$\\
\hline \hline
 \label{tab:KsuKsd}
\end{tabular}
%\end{minipage}
\end{table*}

We also stress that in the free-fall supersonic accretion case which
is realized for high X-ray luminosities ($L_x>4\times 10^{36}$~erg/s),
spin evolution of a NS is entirely determined by the sign of the captured
specific angular momentum, and, strictly speaking, the equilibrium
period cannot be determined from the mass accretion rate and
NS magnetic field only; in real systems the equilibrium state
still could be established due to alternating spin-up/spin-down torques in the
inhomogeneous stellar wind, but this cannot be taken into account in
our simulations. For prograde angular momentum accretion, the NS in
this regime will spin-up until the corotation radius will become
comparable with the magnetospheric one, $R_{\rm A}=R_{\rm c}$, most
likely resulting in the transient accretion or propeller regime. Transient sources
are beyond the scope of the present paper.

The values of spin-up and spin-down torques exerted on a rotating
magnetized NS are summarized in Table \ref{tab:KsuKsd}. The spin
evolution of the NS with these torques at each state can be
explicitly understood as a tendency to reach an equilibrium period,
$P_{\rm eq}$, which is derived from the condition $K=0$. Equilibrium
periods are also listed in Table  \ref{tab:KsuKsd}.

As mentioned in \S \ref{sec:mstr},  mass loss and  and mass exchange
can change the orbital period. In the ejector and propeller states,
the matter transferred from the companion is ejected out of binaries.
In this work, the ejection of matter is treated as occurring in Jeans mode, i.e.
as spherical wind ejection with the specific angular momentum equal
to the NS orbital one.

Since overwhelming majority of NS in SyXB form via core-collapses,
we set initial spin periods of  neutron stars equal to 10\,ms\
\citep{Cordes2004}.

\subsubsection{Magnetic field evolution}
\label{sec:mfe} For the decay of the magnetic field of an accreting NS,
 a commonly accepted idea does not exist as yet.
\cite{Bisnovatyi-Kogan1974} suggested that the decay results from
the screening of the original magnetic field by the accreted matter.
\cite{Ruderman1991} suggested that the decay is due to   the
crustal motion on the surface of  NS. \cite{Goldreich1992}
considered decrease of the magnetic field due to Ohmic decay.

From the statistical analysis of 24 binary radio pulsars,
\cite{Heuvel1995} discovered a clear correlation between the
magnetic field and the mass accreted by the NS. In our study, we
assume that magnetic field depends exponentially on  the amount of
accreted matter and use the formula suggested by \cite{Oslowski2011}:
\begin{equation}
B_{\rm NS}=(B^{\rm i}_{\rm NS}-B_{\rm min})\exp(-\frac{\Delta
M}{M_{\rm B}})+B_{\rm min},
\label{eq:mag}
\end{equation}
where $\Delta M$ is accreted mass, $M_{\rm B}$ is magnetic decay
mass scale and $B_{\rm min}$ is the minimal magnetic field of NS.
 We set $M_{\rm B}=0.025 M_\odot$ and $B_{\rm min}=10^8$
Gauss. $B_{\rm NS}^{\rm i}$ is initial magnetic field of NS. We
assume that the initial magnetic fields of  nascent NS  are
 distributed log-normally \citep{FGK2006,Popov_ea2010}. The median value and the dispersion of
$\log B_{\rm NS}^{\rm i}$ are 12 and 1, respectively.

The decay of the magnetic field of a non-accreting NS should be
similar to that of an isolated NS.  We assume, after
\cite{Kiel2008}, that magnetic field decays exponentially due to
Ohmic decay
\begin{equation}
B_{\rm NS}=B_{\rm NS}^{\rm i}{\exp}(-\frac{t-t_{\rm acc}}{\tau_{\rm
B}}),  \label{eq:smb}
\end{equation}
where $t$ is the age of the NS, $\tau_{\rm B}$ is magnetic field
decay time scale and $t_{\rm acc}$ is the time which NS has spent
accreting matter via Roche lobe overflow. \cite{Kiel2008} have shown
that $\tau_{\rm B}=2$ Gyr is optimal in their preferred models of
Galactic pulsar population. We use this value of  $\tau_{\rm B}$ in
our model.

\section{An example of  NS evolution}

\begin{figure*}
\begin{tabular}{c@{\hspace{3pc}}c}
\includegraphics[totalheight=2.5in,width=2.0in,angle=-90]{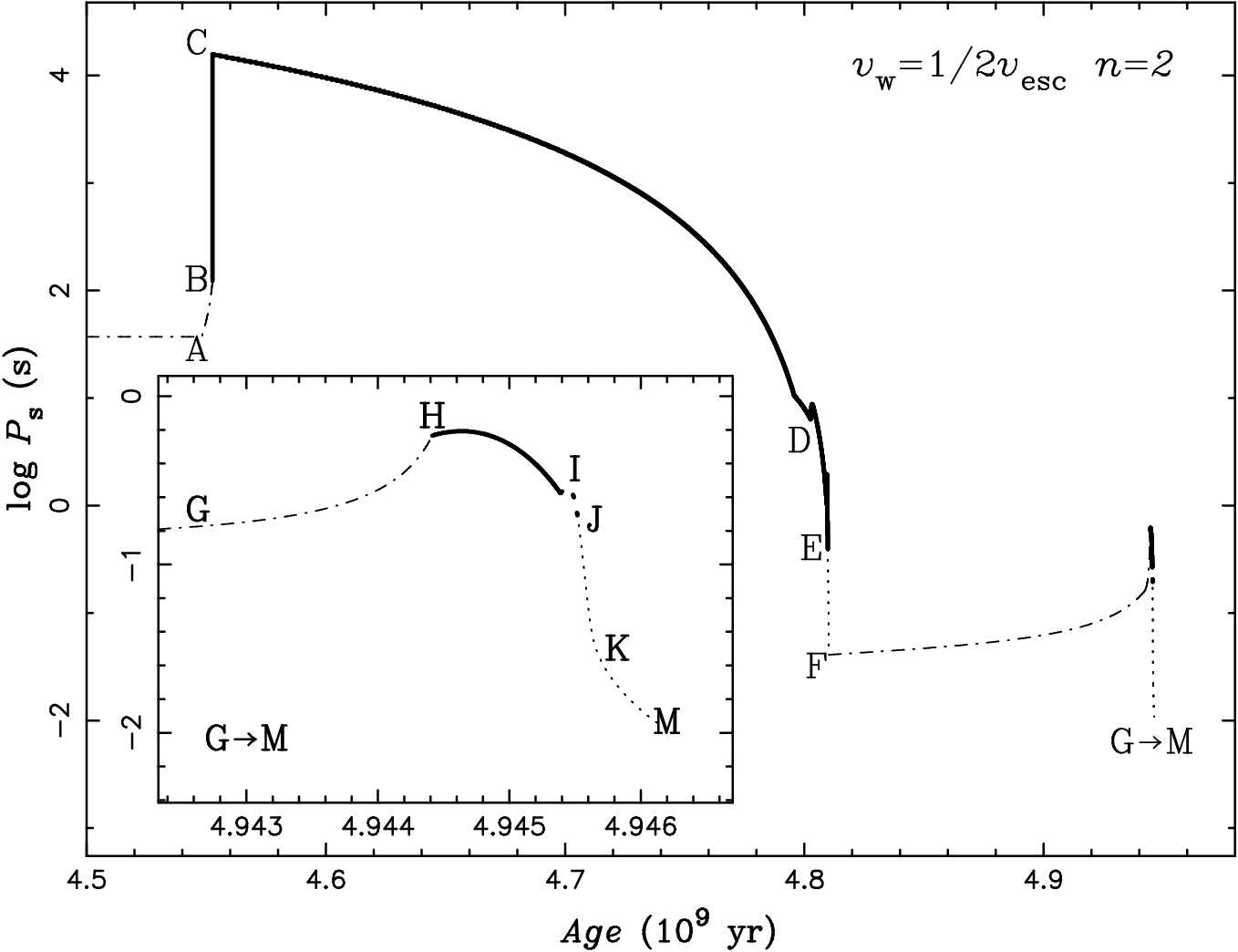}&
\includegraphics[totalheight=2.5in,width=2.0in,angle=-90]{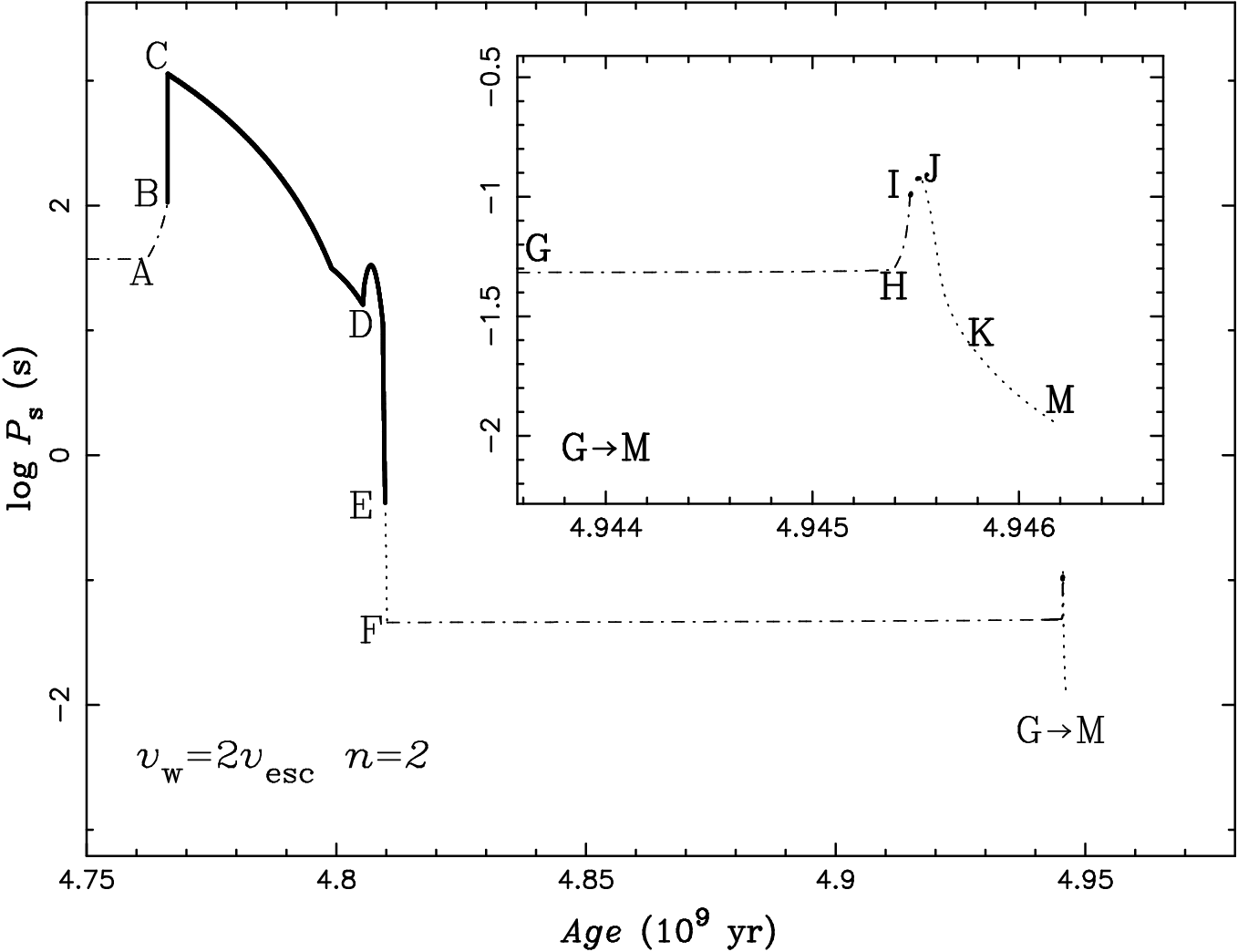}\\
\includegraphics[totalheight=2.5in,width=2.0in,angle=-90]{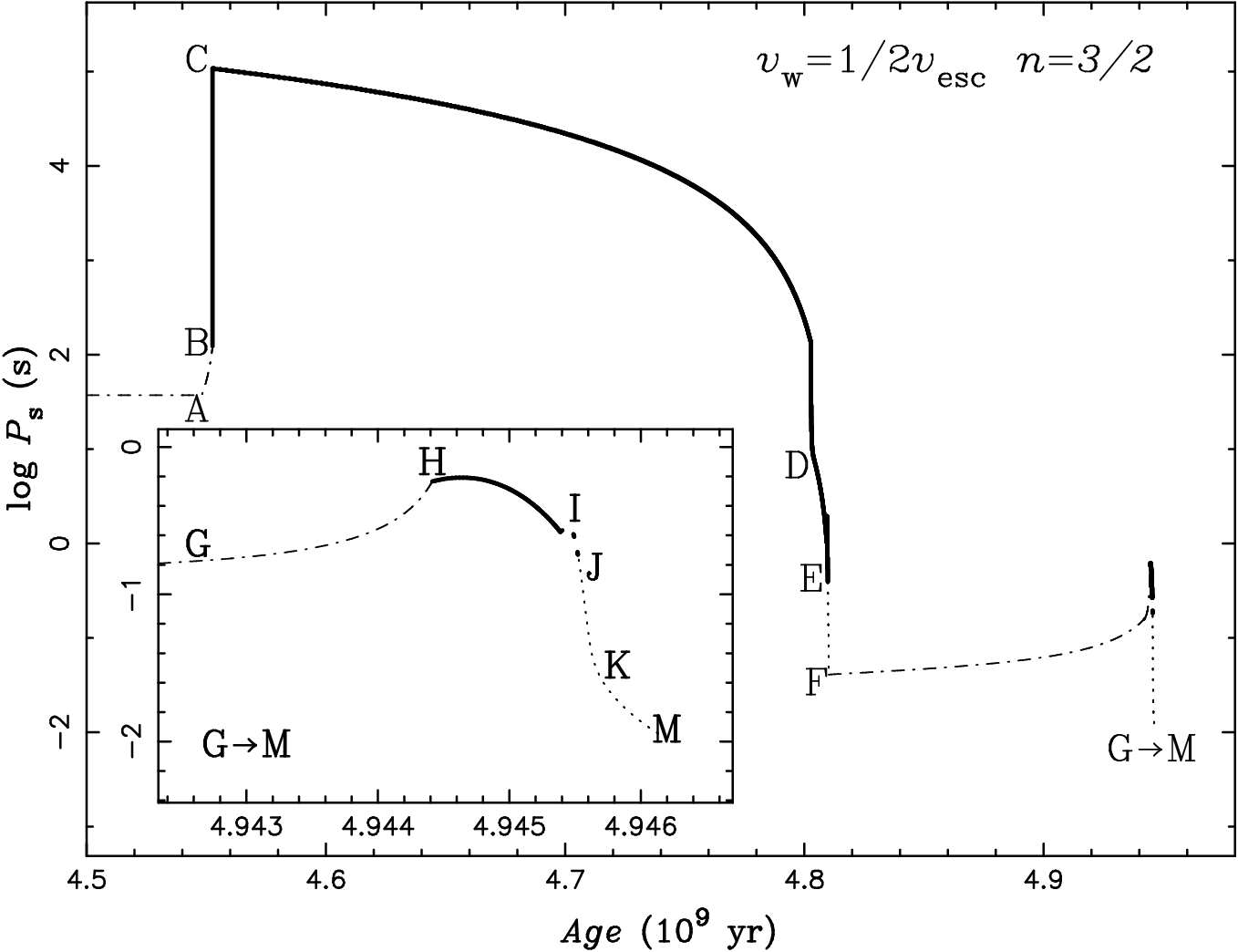}&
\includegraphics[totalheight=2.5in,width=2.0in,angle=-90]{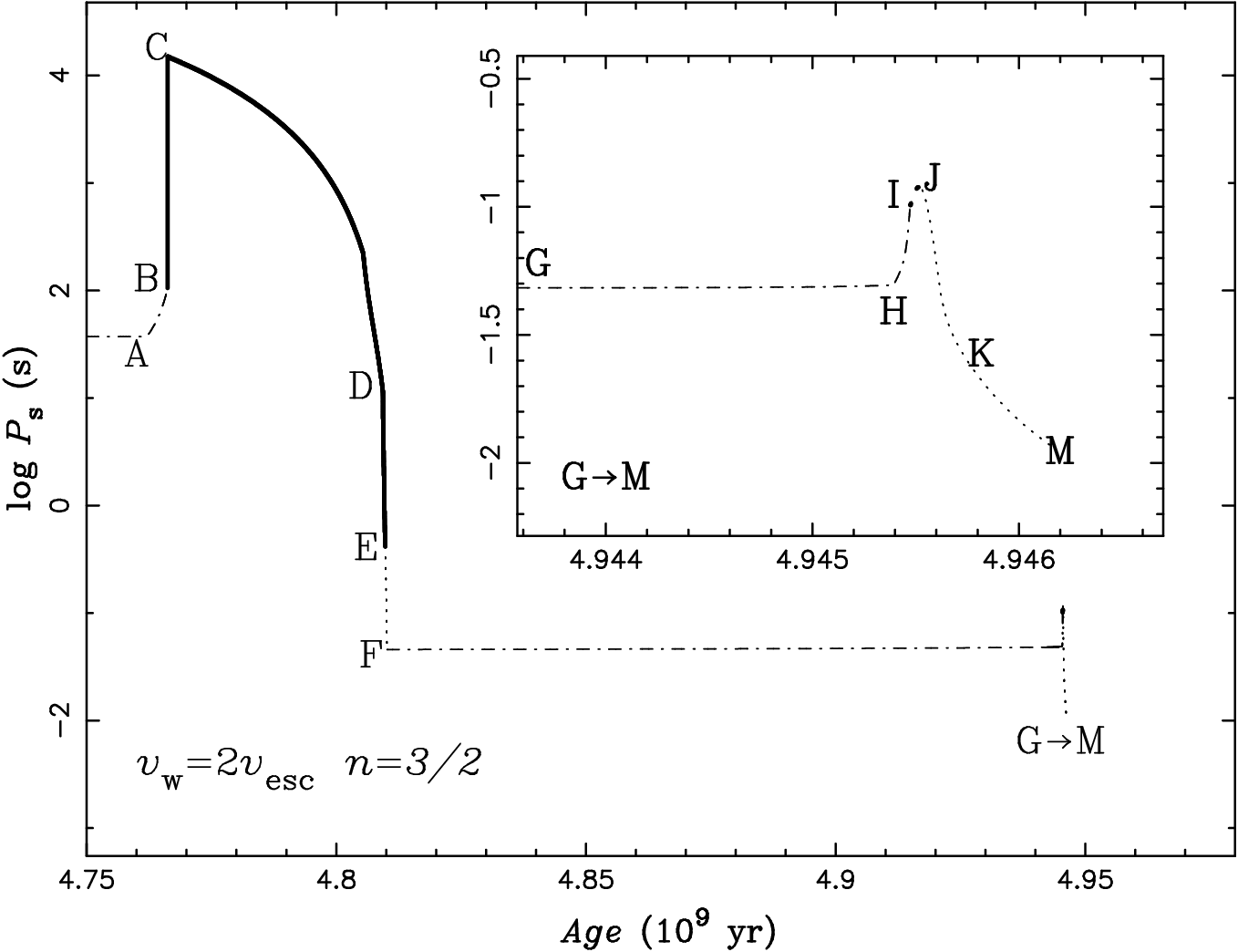}\\
\end{tabular}
\caption{---Spin evolution of an accreting magnetized NS in a binary
system for different input parameters.  Light dot-dashed lines
represent the states in which NS evolves like an isolated NS or it
is in the propeller state, that is, the binary does not emit X-rays.
Thick solid lines represent the states in which the binary is a
SyXB. Thick and thin dotted lines correspond to the states in which the
binary is a LMXB and mass accretion rate onto the NS is lower (between
points 'I' and 'J') or higher (between  'J' and 'M') than the
Eddington critical accretion rate, respectively. In the left two
panels, the lines from  'B' to 'E' and from  'H'
to 'I' represent the states in which the binary is a SyXB. In the
right two panels, the binary is a SyXB
between points 'B' and 'E'. The inserts in the panels
show detailed evolution between points 'G' and 'M'. See the text and
Table \ref{tab:ev} for more details.}

\label{fig:ev}
\end{figure*}

\begin{table*}
  \caption{
Some important parameters in the turning points of spin evolution
marked by the letters in the top left and
 bottom right panels of Fig.~\ref{fig:ev}.
The 2nd column gives  the evolutionary age of the secondary. Columns
3 and 4 give the masses of the NS and the secondary, respectively.
The letters in the parentheses in column 4 indicate the evolutionary
stage of the secondary: FGB, BHeC, E-AGB and AGB stay for the first
giant branch (FGB), core helium burning (BHeC), early asymptotic
giant branch (E-AGB), thermally-pulsing asymptotic giant branch
(TP-AGB). The radius of the secondary is given in column 5. Columns
6 and 7 list the terminal stellar wind velocity and the
mass-accretion rate, respectively. When the secondary overfills
Roche lobe, the wind does not play any role. Columns  8, 9, and 10
give the gravitational capture radius $R_{\rm G}$,
 the magnetospheric Alfv\'{e}n radius $R_{\rm A}$ and corotation
radius $R_{\rm c}$, respectively. Column 11 gives the total torque.
Columns 12 and 13 show magnetic field and spin period.}
  \tabcolsep0.90mm
  \begin{tabular}{ccccccccccccc}
  \hline\hline
&&&&&$v_{\rm w}=1/2v_{\rm esc}$&$n=2$&&&&&\\
Point&Age&$M_{\rm NS}$&$M_{\rm 2}$&$R_{\rm 2}$&$v_{\rm
w}$&$\dot{M}_{\rm NS}$&$R_{\rm G}$&$R_{\rm A}$ or $R_{\rm
SA}$&$R_{\rm c}$&$K_{\rm su}-K_{\rm sd}$&$B_{\rm NS}$&$P_{\rm
s}$\\
&($10^6$yr)&$ (M_\odot)$&$ (M_\odot)$&$ (R_\odot)$&(km
s$^{-1}$)&$(10^{16}{\rm g\
s^{-1}})$&(cm)&(cm)&(cm)&(dyn cm)&({\rm G})&(s)\\

A&4548.07&1.40&1.30(FGB)&3.26&194.9&$6.5\times10^{-6}$&$2.4\times10^{12}$&
$4.2\times10^{10}$&$1.9\times10^{10}$&$-9.0\times10^{29}$&$5.1\times10^{11}$
&37.18\\
B&4552.66&1.40&1.30(FGB)&3.28&194.3&$6.6\times10^{-6}$&$2.4\times10^{12}$&
$4.2\times10^{10}$&$4.2\times10^{10}$&$-9.2\times10^{29}$&$5.1\times10^{11}$
&123.1\\
C&4552.67&1.40&1.30(FGB)&3.28&194.3&$6.7\times10^{-6}$&$2.4\times10^{12}$&
$4.2\times10^{10}$&$4.2\times10^{10}$&$6.6\times10^{30}$&$5.1\times10^{11}$
&15760\\
D&4802.68&1.40&1.28(FGB)&38.5&56.4&$3.6\times10^{-1}$&$2.6\times10^{12}$&
$4.9\times10^{10}$&$5.7\times10^{10}$&$-6.5\times10^{31}$&$5.1\times10^{11}$
&6.328\\
E&4809.78&1.41&1.21(FGB)&136&---&$1.6\times10^{2}$&---&
$2.5\times10^{7}$&$9.0\times10^{7}$&$0.0$&$3.9\times10^{11}$
&0.3927\\
F&4810.18&1.44&1.16(BHeC)&12.2&94.9&$4.5\times10^{-3}$&$1.0\times10^{12}$&
$5.1\times10^{8}$&$2.0\times10^{7}$&$-1.7\times10^{31}$&$9.7\times10^{10}$
&0.04093\\
G&4941.68&1.44&1.12(E-AGB)&24.5&66.2&$8.7\times10^{-2}$&$2.0\times10^{12}$&
$3.0\times10^{8}$&$4.9\times10^{7}$&$-8.8\times10^{31}$&$9.7\times10^{10}$
&0.1551\\
H&4944.42&1.44&1.11(E-AGB)&53.1&44.8&$1.9$&$4.0\times10^{12}$&
$1.2\times10^{8}$&$1.2\times10^{8}$&$-2.0\times10^{32}$&$9.7\times10^{10}$
&0.5839\\
I&4945.39&1.44&1.10(E-AGB)&112&---&$4.9$&---&
$5.2\times10^{8}$&$7.0\times10^{8}$&$0.0$&$9.2\times10^{10}$
&0.2656\\
J&4945.53&1.44&1.10(E-AGB)&136&---&$1.8\times10^{2}$&---&
$1.3\times10^{7}$&$5.3\times10^{7}$&$0.0$&$8.9\times10^{10}$
&0.1723\\
K&4945.77&1.47&1.00(TP-AGB)&219&---&$5.7\times10^{3}$&---&
$8.9\times10^{6}$&$1.3\times10^{7}$&$0.0$&$3.7\times10^{10}$
&0.0216\\
M&4946.17&1.49&0.55(TP-AGB)&251&---&$2.5\times10^{3}$&---&
$5.4\times10^{6}$&$8.4\times10^{6}$&$0.0$&$1.2\times10^{10}$
&0.001085\\

%%%%%%%%%%%%%%%%%%%%%%%%%%%%%%%%%%%%%%%%%%%%5555
%%%%%%%%%%%%%%%%%%%%%%%%%%%%%%%%%%%%%%%%%%%%%%%%
&&&&&$v_{\rm w}=2v_{\rm esc}$&$n=3/2$&&&&&\\
Point&Age&$M_{\rm NS}$&$M_{\rm 2}$&$R_{\rm 2}$&$v_{\rm
w}$&$\dot{M}_{\rm NS}$&$R_{\rm G}$&$R_{\rm A}$ or $R_{\rm
SA}$&$R_{\rm c}$&$K_{\rm su}-K_{\rm sd}$&$B_{\rm NS}$&$P_{\rm
s}$\\
&($10^6$yr)&$ (M_\odot)$&$ (M_\odot)$&$ (R_\odot)$&(km
s$^{-1}$)&$(10^{16}{\rm g\
s^{-1}})$&(cm)&(cm)&(cm)&(dyn cm)&({\rm G})&(s)\\

A&4762.05&1.40&1.30(FGB)&10.4&436.0&$6.4\times10^{-6}$&$7.6\times10^{11}$&
$4.0\times10^{9}$&$1.9\times10^{9}$&$-8.6\times10^{29}$&$4.6\times10^{11}$
&37.21\\
B&4766.22&1.40&1.30(FGB)&11.0&423.2&$8.4\times10^{-6}$&$8.0\times10^{11}$&
$3.8\times10^{9}$&$3.8\times10^{9}$&$-1.0\times10^{30}$&$4.6\times10^{11}$
&105.5\\
C&4766.23&1.40&1.30(FGB)&11.0&423.2&$8.4\times10^{-6}$&$8.0\times10^{11}$&
$3.8\times10^{9}$&$3.8\times10^{9}$&$6.8\times10^{30}$&$4.6\times10^{11}$
&15020\\
D&4805.38&1.40&1.27(FGB)&51.2&194.8&$9.0\times10^{-3}$&$3.3\times10^{12}$&
$1.3\times10^{9}$&$6.2\times10^{9}$&$1.2\times10^{30}$&$4.6\times10^{11}$
&22.47\\
%E&4809.36&1.40&1.23(FGB)&---&$3.9$&---&
%$1.4\times10^{8}$&$8.5\times10^{8}$&$0.0$&$4.6\times10^{11}$
%&11.3\\
E&4809.80&1.40&1.21(FGB)&115&---&$6.4\times10^1$&---&
$2.6\times10^{7}$&$9.3\times10^{7}$&$0.0$&$4.1\times10^{11}$
&0.4146\\
F&4810.18&1.44&1.16(BHeC)&12.2&380.0&$2.1\times10^{-5}$&$1.0\times10^{12}$&
$1.5\times10^{9}$&$2.2\times10^{7}$&$-9.9\times10^{29}$&$1.1\times10^{11}$
&0.04572\\
G&4941.68&1.44&1.13(E-AGB)&24.5&265.0&$5.0\times10^{-4}$&$2.0\times10^{12}$&
$8.3\times10^{8}$&$2.2\times10^{7}$&$-5.6\times10^{30}$&$1.1\times10^{11}$
&0.04807\\
H&4945.39&1.44&1.11(E-AGB)&113&---&$4.9$&---&
$5.8\times10^{7}$&$2.3\times10^{7}$&$-1.6\times10^{34}$&$1.1\times10^{11}$
&0.04954\\
I&4945.47&1.44&1.11(E-AGB)&124&---&$2.5\times10^1$&---&
$3.7\times10^{7}$&$3.7\times10^{7}$&$-6.4\times10^{34}$&$1.1\times10^{11}$
&0.1016\\
J&4945.53&1.44&1.10(E-AGB)&137&---&$1.8\times10^{2}$&---&
$1.4\times10^{7}$&$4.0\times10^{7}$&$0.0$&$1.1\times10^{11}$
&0.1164\\
K&4945.77&1.46&1.00(TP-AGB)&218&---&$5.5\times10^{3}$&---&
$9.6\times10^{6}$&$1.4\times10^{7}$&$0.0$&$4.5\times10^{10}$
&0.02403\\
M&4946.17&1.49&0.55(TP-AGB)&251&---&$2.5\times10^{3}$&---&
$5.7\times10^{6}$&$8.7\times10^{6}$&$0.0$&$1.4\times10^{10}$
&0.01137\\

 \hline \hline
 \label{tab:ev}
\end{tabular}
%\end{minipage}
\end{table*}

Symbiotic X-ray binaries are wide and faint X-ray
systems \citep{Masetti2006}. In this work, binaries composed of a
 NS and a giant or a
giant-like star are considered as SyXBs if they
satisfy the following conditions:\\
(i)The systems are detached;\\
(ii)NSs are in the accretor state.\\
\cite{Revnivtsev2011} suggested that some LMXBs have giant or
giant-like donors which fill their Roche lobes, such as GX 13+1 and
Cyg X-2. Usually, these giant donors are still very close to the main
sequence and have small He cores. In these LMXBs, the X-ray
luminosities are very high. Compared to SyXBs, their orbital
periods are very short. In order to understand SyXBs, we also
discuss these LMXBs. Further, we use notation ``LMXB'' for the  systems in which
giant or giant-like donors fill their  Roche lobes.

Below, we present a detailed example of the evolution of an
accreting magnetized NS in a binary system, as it emerges using the
algorithm for spin and magnetic field evolution presented in the previous
sections. We start with a binary which consists of a NS and a
main-sequence star. At the onset of evolution, the mass of NS is 1.4
$M_\odot$, its surface magnetic field strength is
 $5\times10^{12}$\,G, and spin period is
0.01\,s. Main-sequence star mass is 1.3 $M_\odot$. Initial orbital
period of the system is 400 day. Figure \ref{fig:ev} shows
evolution of the NS spin period for different input parameters. In
Table~\ref{tab:ev} we present some important physical values for all
points marked by letters in the top left ($v_{\rm w}=1/2v_{\rm esc}$
and $n=2$) and bottom right ($v_{\rm w}=2v_{\rm esc}$ and $n=3/2$)
panels in Fig.~\ref{fig:ev}. For $v_{\rm w}=1/2v_{\rm esc}$ and
$n=2$,
the evolution of $P_{\rm s}$ proceeds as follows.\\

Before point A, the secondary is in the first giant branch (FGB)
stage, but does not fill  its Roche lobe. Bondi accretion rate of
the NS is below $10^{-15} M_\odot$ yr$^{-1}$. Neutron  star  evolves
like an isolated object. At point A, mass-capture rate  by NS
exceeds $10^{-15} M_\odot$ yr$^{-1}$\ and  NS starts interacting
with its environment.
 At point A, relations $R_{\rm G}>R_{\rm A}>R_{\rm c}$ (See Table \ref{tab:ev})  and $j_{\rm
a}<j_{\rm K}(R_{\rm A})$ hold and  NS enters the propeller state of
quasi-spherical accretion. While NS evolves through the propeller
state from point A to point  B, its spin period increases and
$R_{\rm c}$ increases. At point B, $R_{\rm A}$ becomes lower than
$R_{\rm c}$, but  $\dot{M}_{\rm NS}<4\times10^{16}$g/s. NS enters
the quasi-spherical subsonic settling accretion state, after which
the evolution of $P_{\rm s}$ is determined by the torque given by
Eqs. (\ref{eq:ksu}) and (\ref{eq:ksd}). The stage of SyXB starts. At
the beginning, $K_{\rm sd}>K_{\rm su}$, and the spin period is
increasing. With the increase of the spin period, $K_{\rm sd}$
decreases. Soon, at point C, $K_{\rm sd}$ $\sim K_{\rm su}$, the
evolution of the spin period is now determined by
Eq.~(\ref{eq:weq}). With the ascend along FGB, stellar-wind velocity
of the giant $v_{\rm w}$ decreases. This results in the increase of
$j_{\rm a}$. At point D, $j_{\rm a}>j_{\rm K}(R_{\rm A})$ and a
wind-fed accretion disk forms around  NS. After point D, the torque
exerted on the NS is given by Eq (\ref{eq:dtor}). Due to large
$P_{\rm s}$ and low $\dot{M}_{\rm NS}$,  immediately after point D
the spin-up torque is smaller than the spin-down torque, and $P_{\rm
s}$ increases from 6.3~s to 8.7~s.
 But very soon, with the enhancement of
$\dot{M}_{\rm NS}$ and $P_{\rm s}$, the spin-up torque becomes
larger than the spin-down torque, and $P_{\rm s}$ starts to
decrease.

At point E the secondary overflows its Roche lobe, the system
becomes LMXB (in our notation). The mass-accretion rate immediately increases and
becomes higher than the Eddington mass-accretion rate. Neutron star
becomes a super-accretor.

At point F, the secondary leaves FGB and evolves to the core
helium burning stage. Mass-loss rate rapidly decreases, but the
velocity of stellar wind is low, so that $j_{\rm a}>j_{\rm K}(R_{\rm
A})$, and  neutron star becomes a propeller with a wind-fed accretion disk.
In this state the mater from the disk does not reach the NS due to
centrifugal barrier.

At point G, the secondary evolves to the early asymptotic giant branch
(E-AGB). With the ascend along AGB,  mass-loss rate of the secondary
increases. Due to the high mass-accretion rate,  spin period $P_{\rm
s}$ rapidly increases.  At point H, $R_{\rm A}\sim R_{\rm c}$ and
$j_{\rm a}>j_{\rm K}(R_{\rm A})$. The system becomes SyXB with a
wind-fed accretion disk. At point I the secondary overflows its
Roche lobe, the system becomes LMXB. Very soon, the NS becomes a
super-accretor at the point J. At point K the secondary evolves into
thermally pulsing asymptotic giant branch star. At point M the
secondary leaves AGB.

To summarize, the system is a SyXB between points 'B' and 'E' and
 'H' and 'I'. It is a LMXB from point 'E' to point 'F' and from
point 'I' to point 'M'.

In our model, in the stages of evolution when the mass-accretion
rate onto NS is higher than $4\times10^{16}$g/s, $v_{\rm w}$ usually
is low, such that $j_{\rm a}>j_{\rm K}(R_{\rm A})$ and a disk forms.
Therefore, the quasi-spherical supersonic accretion hardly occurs in
SyXBs.

The case of high wind  velocity ($v_{\rm w}=2v_{\rm esc}$) and
respective spin period evolution are shown in the right panels of
Figure~\ref{fig:ev}. High velocity of the stellar wind from the
companion to NS results in a low mass-accretion rate by the latter,
which gives a large $R_{\rm A}$. As Figure \ref{fig:ev} and Table
\ref{tab:ev} show for the simulation with $v_{\rm w}=2v_{\rm esc}$,
at point B of the track when the binary becomes SyXB, the age of the
optical star is larger  than the one in the simulation with $v_{\rm
w}=1/2v_{\rm esc}$. In the $v_{\rm w}=2v_{\rm esc}$ case,    the
binary never becomes a wind-fed SyXB when the donor is in the  AGB
stage, but it may become a LMXB for a short time.

In general, $R_{\rm A} < R_{\rm G}$ in the accretor state. The
spin-up torques in the case of quasi-Keplerian angular momentum distribution
in the hot shell at the subsonic accretion state
($n=3/2$, bottom panels  in Fig.~\ref{fig:ev}), are smaller than
those in the case of iso-angular momentum distribution ($n=2$, top
panels in Fig.~\ref{fig:ev}). Neutron stars in the bottom panels
have longer spin periods at point C. Therefore, the low value of
$n$ is favourable to explain
SyXBs in which NS have long spin periods.

\section{Population of symbiotic X-ray binaries}

As we mentioned above, theoretical models of the population of LMXBs
depend on badly known input parameters, such as kick velocity and
common envelope treatment \citep[e. g.,
][]{1995ApJS..100..233I,Pfahl2003,Zhu2012}. However, in this
pioneering study of SyXBs we focus on the effects which are
important for the observational appearance of the latter: the donor
wind velocity ($v_{\rm w}$) and  the angular momentum distribution
in the shell of matter settling onto NS (index $n$).

\subsection{The Number and Birthrate of the Galactic SyXBs}
\label{sec:num}

About 70\% (for $v_{\rm w}=1/2v_{\rm esc}$) --- 98\% (for $v_{\rm w}=2v_{\rm
esc}$) of all SyXBs NS are formed via core-collapse, while other
$\simeq30$\% ($v_{\rm w}=1/2v_{\rm esc}$)--- 2\% ($v_{\rm w}=2v_{\rm
esc}$) via ECS. The reason for negligibly low contribution of
post-AIC systems is large initial separation of components which is
necessary for formation of an ONeMg WD. The initial orbital periods
%$P_{\rm s}^{\rm i}$
of the progenitors of SyXBs in which an AIC may happen
 are the longest.
After formation of the WD, the orbit has to remain  wide, so that
the secondary can evolve to the red giant stage (this means that the
efficiency of the common envelope ejection must be high). If the
orbit is wide, the accretion rate is low and unfavourable for AIC.
In the systems which produce NS via ECS the primaries usually have
ZAMS mass between $\sim 8$ and 11 $M_\odot$ and short initial
orbital periods. They encounter a common envelope phase when the
primaries evolve through HG or FGB. The primaries become naked
helium stars and finally form NS. Core-collapse SNe occur in massive
stars ($ \ge 11 M_\odot$) and for progenitors of SyXBs with
initially massive primaries the range of initial orbital periods is
limited by the possibility of the system to remain bound after SN
explosion.

The input parameter $n$, describing the angular momentum
distribution in the hot shell above the NS magnetosphere at the
quasi-spherical accretion settling stage, is found to have a
negligible effect on SyXBs' birthrate and number. However, the
terminal wind velocity $v_{\rm w}$  greatly influences the birthrate
and number of SyXBs and introduces an uncertainty of up to a factor
of about 10. The Galactic birthrate and number of SyXB in the
simulation with low terminal wind velocity, $v_{\rm w}=1/2v_{\rm
esc}$ are  $\simeq 4.1\times 10^{-5}$ yr$^{-1}$ and $\simeq 1000$;
in the simulation with high terminal wind velocity $v_{\rm
w}=2v_{\rm esc}$ they are lower, $\sim 6.6\times 10^{-6}$ yr$^{-1}$
and $\sim 100$. Up to now, only 10 SyXBs or candidates were
discovered. A possible reason is that SyXBs are difficult to find
because of their low X-ray luminosities and transient character of
wind accretion.

In contrast to SyXBs,
the input parameters $n$ and $v_{\rm w}$ are found to have
a negligible effect on LMXB (semidetached systems in our notation).
population. In the simulation with  $v_{\rm w}=1/2v_{\rm esc}$, the
Galactic birthrate of LMXBs is $\sim 7.0\times 10^{-5}$ yr$^{-1}$,
and the number is $\sim 10000$. In about 50\% of all simulated LMXBs, NS was formed
via core-collapse, while in another $\simeq50$\% -- via ECS.
The contribution of post-AIC systems is
negligible. Taken at face value, these numbers suggest that LMXBs
with giant donors comprise a significant fraction of Galactic LMXBs
(cf. \citet{Pfahl2003}). This also means that we encounter the known
problem of ``overproduction'' of LMXBs: while population synthesis
codes produce $\sim 10^3 - 10^4$\ strong X-ray systems, the number
of observed ones is less than 200 \citep{Liu2006}. \textit{ Ad hoc }
solution, suggested by \citet{Pfahl2003} is possible cyclic
behaviour of most LMXB due to irradiation effects, though,
self-consistent solution is not available, as yet. However,
\citet{Buning2004} found that irradiation does not affect giants.
The problem remains open.

\subsection{Properties of SyXBs}

\begin{figure}
\includegraphics[totalheight=3in,width=2.5in,angle=-90]{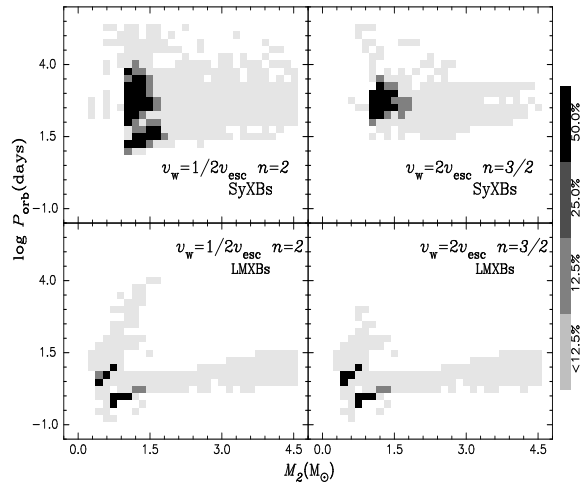}
\caption{--- Distributions of the orbital periods $P_{\rm orb}$ of
             SyXB and LMXB vs. masses of their secondary stars.
             Two upper panels are for the SyXBs population,
            two lower panels for LMXBs population.
Gradations of gray-scale correspond to the
number density of systems $>$1/2,
            1/2 -- 1/4, 1/4 -- 1/8, 1/8 -- 0 of the maximum of
             ${{{\partial^2{N}}\over{\partial {\log P_{\rm orb}}{\partial {\log
            M_2}}}}}$ and blank regions do not contain any stars.
            }
\label{fig:mp}
\end{figure}

Figure \ref{fig:mp} presents, in gray-scale, distributions of the
orbital periods $P_{\rm orb}$ of SyXBs and LMXBs vs. masses of their
secondary components $M_{\rm 2}$. Two upper panels show the
population of SyXBs, while two lower panels present LMXBs. The
orbital periods of model SyXBs exceed $\simeq$25~days and extend to $\simeq$30000~days.
The range of orbital periods of the known SyXBs is from 50.2~day (IGR J16393-4643)
to 1161~day (GX 1+4).  In the distribution for the LMXB
population there are two peaks of $M_2$ and $P_{\rm orb}$. The upper
left peak is due to the systems with long periods in which the
secondaries can evolve to the FGB stage and even AGB stage. The peak
located lower and to the right is formed mainly by LMXBs in which
the secondaries are in the Hertzsprung gap and overflow Roche lobe
because of short orbital periods. Comparison of the left and the
right panels shows that high wind velocity is unfavourable for
formation of the SyXBs and LMXBs with long orbital periods. As
Figure~\ref{fig:mp} shows,  SyXBs are predominantly the systems
with large $M_2$ and $P_{\rm orb}$.

\begin{figure}
\includegraphics[totalheight=3.3in,width=3.3in,angle=-90]{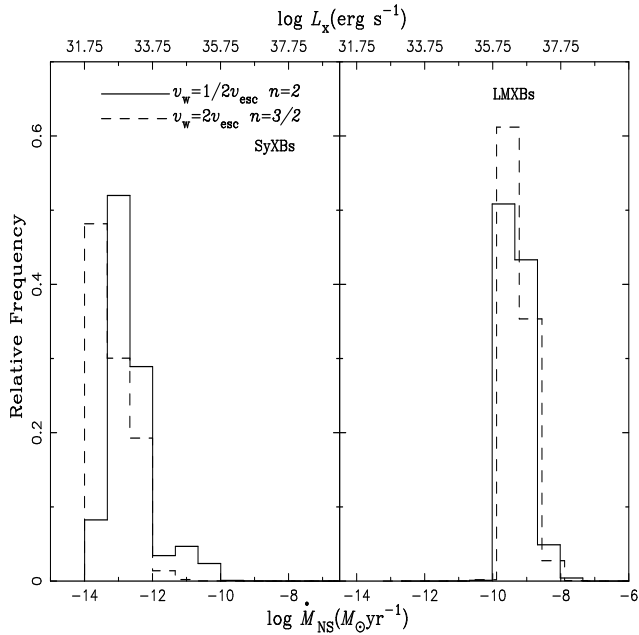}
\caption{---Distribution of accretion rates (X-ray luminosities) by
NSs in SyXBs and LMXBs.}\label{fig:ma}
\end{figure}

Figure \ref{fig:ma} shows the distribution of accretion rates (X-ray
luminosities) of NS in SyXBs and LMXB. X-ray luminosities
are approximated as
\begin{equation}
\begin{array}{l}
L_{\rm x} = \eta \dot{M}{\rm c}^2 \simeq 5.7\times10^{35}{\rm erg \
s^{-1}}(\frac{\eta}{0.1})\times(\frac{\dot{M}}{{10^{-10}{ M_\odot
{\rm yr^{-1}}}}}),
\end{array}\label{eq:lx}
\end{equation}
where $\eta \simeq 0.1$ is the efficiency of converting accreted
mass into X-ray photons. The peak in the distribution of
mass-accretion rates occurs at $\sim 10^{-13} M_\odot$ yr$^{-1}$,
and their X-ray luminosities are between $10^{32}$ and $10^{36}$ erg
s$^{-1}$. Therefore, SyXBs are faint X-ray sources.
 Our results are consistent with observations. LMXBs
have high mass-accretion rates($\sim 10^{-9} M_\odot$ yr$^{-1}$). In
our work, about 1\% of LMXBs have super-Eddington accretion.

\begin{figure}
\includegraphics[totalheight=3.3in,width=3.3in,angle=-90]{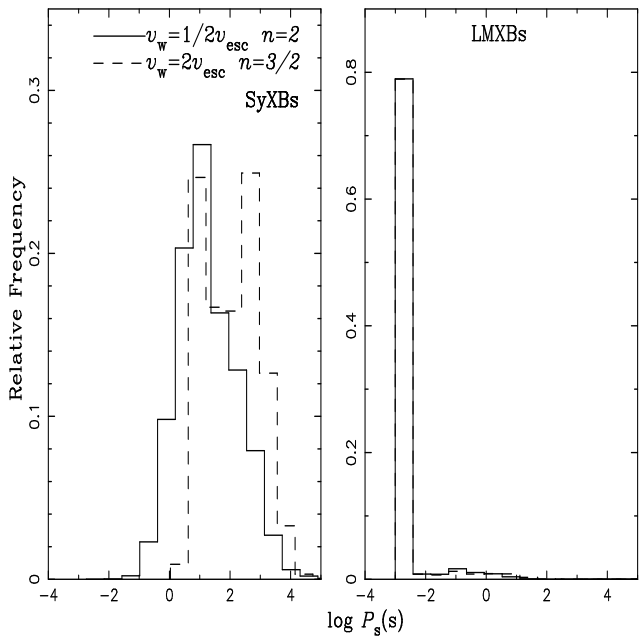}
\caption{--- Number distribution of the spin periods of NSs in SyXBs
and LMXBs.  The left panel is for population of SyXBs, the right
panel is for LMXBs. The numbers are normalized to 1.
}\label{fig:spin}
\end{figure}

Figure \ref{fig:spin} shows the distributions of spin periods. The
range of spin periods in SyXBs is between 0.1 s and $\sim 10^5$ s.
In the simulation with $v_{\rm w}=2v_{\rm esc}$ and $n=3/2$, there
are two peaks. The left peak is formed by SyXBs with wind-fed
accretion disks, while the right peak is due to quasi-spherically
accreting SyXBs because  small $n$ is favourable for producing  long
spin periods. In this simulation, about 20\% of SyXBs have wind-fed
accretion disks. However, in the simulation with $v_{\rm w}=1/2v_{\rm
esc}$ and $n=2$, only about 7\% of SyXBs have wind-fed accretion
disks. The low wind velocity results in a high mass-accretion rate
which produces shorter spin periods. As the left panel of
Fig.~\ref{fig:spin} shows, there is only  one peak which results mainly
from quasi-spherically accreting SyXBs. Most of LMXBs have very
short spin periods ($\sim 10^{-3}$ s) because NSs have accreted a
large amount of matter from their companions. Due to difference in
amounts of accreted matter, SyXBs have
long spin periods, while LMXBs have very short spin periods. The
latter may be good progenitors of binary millisecond pulsars.

\begin{figure}
\includegraphics[totalheight=3.3in,width=3.3in,angle=-90]{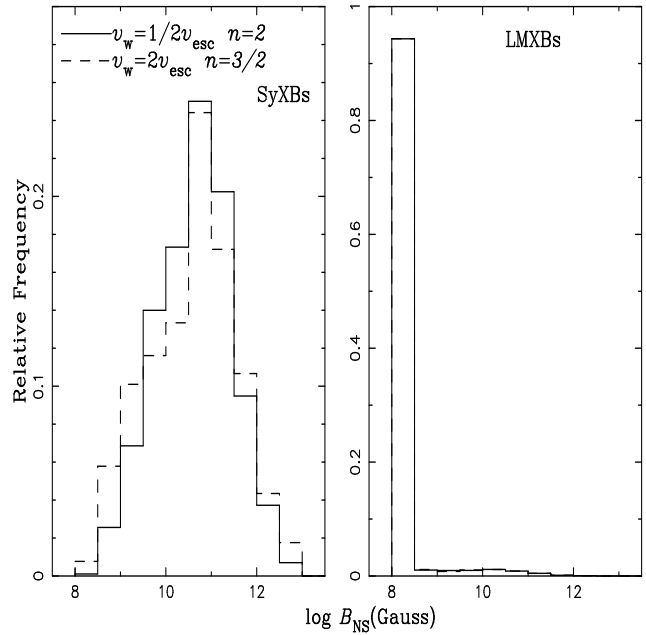}
\caption{---Similar to Figure \ref{fig:spin}, but for magnetic
fields of NSs in SyXBs and LMXBs.}\label{fig:magb}
\end{figure}

Figure \ref{fig:magb} shows the distribution of  magnetic fields of
NSs in SyXBs and LMXBs. The peak of distribution in SyXBs is at
$10^{11}$ G. The magnetic fields of NSs in SyXBs are not
violently suppressed. Most of LMXBs have very weak magnetic fields
($\sim 10^{8}$ G) because NSs have accreted enough matter to
suppress their fields.

\subsection{Individual SyXBs}
\label{sec:indi}

\begin{figure}
\begin{tabular}{c}
\includegraphics[totalheight=3in,width=2.5in,angle=-90]{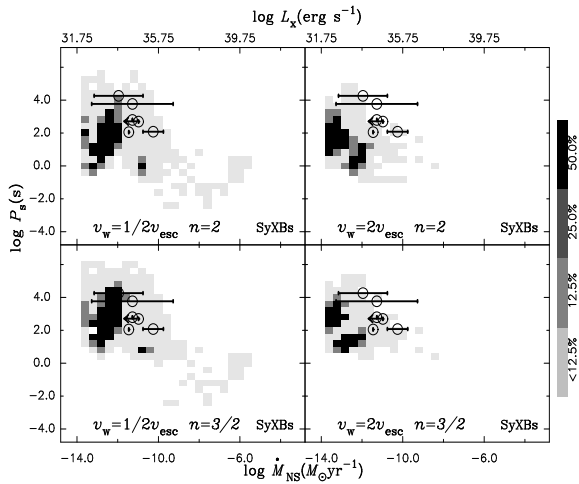}
\end{tabular}
\caption{---Gray-scale maps of the distributions of
            the spin period $P_{\rm s}$
            vs. mass-accretion rate $\dot{M}_{\rm
            NS}$ (or X-ray luminosity $L_{\rm x}$) for SyXBs.
            The data on observed SyXBs from Table \ref{tab:syxb} is
            plotted by cycles.
            }
\label{fig:pp}
\end{figure}

As Table \ref{tab:syxb} shows, spin periods, orbital periods and
X-ray luminosities are the most important parameters of SyXBs. X-ray
luminosities (mass-accretion rates) and spin periods of five SyXBs
observed and our results for the sample of SyXBs and LMXBs are
plotted in Figure \ref{fig:pp}. The models with a low wind velocity
($v_{\rm w}=1/2v_{\rm esc}$) are preferred for explaining the known
SyXBs.

Our results in the simulation with $v_{\rm w}=1/2v_{\rm esc}$ cover
the positions of Sct~X-1,  IRXS~J180431.1-273932 and 2XMM
J174016.0-290337 very well. Our model cannot identify whether there
are accretion disks in these systems. Formation of the disks greatly
depends on the wind velocity.

Neutron stars  in 4U 1954+31 and IGR J16358-4724 have very long
$P_{\rm s}$. In order to explain them by the model of a NS with an
accretion disk, \citet{Patel2007} suggested that IGR J16358-4724
contains a magnetar. Indeed, as Eqs. (\ref{eq:deq}), (\ref{eq:seq}), and
(\ref{eq:weq}) show,  the equilibrium NS spin period $P_{\rm s}$
increases with NS magnetic field. However, the position of
4U~1954+31 and IGR~J16358-4724 in
Fig.~\ref{fig:pp}
fit well the region occupied by quasi-spherically accreting SyXBs
with low wind velocity. Then an ultra-strong magnetic field is not a
necessary condition to explain 4U~1954+31 and IGR~J16358-4724. The
same conclusion was obtained by \cite{Marcu2011} who interpreted the
18300~s pulse period of 4U~1954+31 within the framework of the
quasi-spherical accretion model developed by \cite{Shakura2011}.

GX 1+4 has a very interesting long-term spin
behaviour\citep{GonzalezGalan2011,Shakura2011}. Based on the
properties of an assumed accretion disk around NS,
\cite{Chakrabarty1997} suggested that magnetic field of NS in GX~1+4
should be ultra-strong ($B_{\rm NS} \sim10^{14}$~G). Recently,
however, based on \citep{Shakura2011} model,
\cite{GonzalezGalan2011} argued that GX~1+4 is not in the
disk-accretion state. Instead, the system which currently shows a
steady spin-down and can have NS magnetic field close to $10^{13}$~G,
definitely is a quasi-spherical wind-accreting source. As
Fig.~\ref{fig:pp} shows, our result supports the
suggestion of \citet{GonzalezGalan2011}.

As seen in Fig. \ref{fig:pp}, the main population of Galactic SyXB
in our population synthesis simulations must have on average
lower X-ray luminosities than
actually observed. This difference is most likely due to still very approximate
treatment of the complicated process of stellar wind accretion onto NS in our model,
which ignores temporal variations of the wind properties from cool giants,
wind clumping, possible effects of the orbital eccentricity, etc.
The settling accretion onto magnetized NSs
at small X-ray luminosities can also be unstable \citep{PKchia}.
However, we stress good correspondence between observed NS spin periods
in SyXBs  and the ones obtained in quasi-spherical accretion model.

\section{Conclusions}

Using simulation of the interaction of  magnetized NSs with their
environment in binary systems, we investigated Galactic population
of SyXBs.  In our simulations, the number of Galactic SyXBs is found
to be $\sim$ 100---1000, and the estimate of their birthrate is
between $\sim 4.1\times 10^{-5}$ yr$^{-1}$ and $\sim 6.6\times
10^{-6}$ yr$^{-1}$. The simulated SyXBs population substantially
depends on the properties of the stellar wind velocity from cool giants,
which is one of the model input parameters.
%{\bf In this paper, we
%do not discuss the effects of other  input parameters of
%binary evolution on SyXB's population.}
% Some parameters (such as kick
%velocity amplitude,  CE treatment et al.) can  affect
%LMXB's population.}
%The input parameter stellar wind velocity has a great
%effect on SyXB population.
In our model, SyXBs have found to have wide
orbital periods $\sim$(10
---10000 days), they are faint X-ray sources
$\sim(10^{32}$---$10^{36}$ erg s$^{-1}$), and have long spin periods
$\sim$(0.1 -- $10^5$)~s.

Our model can explain certain observational properties of some known
SyXBs or candidate systems. 4U~1954+31 and IGR~J16358-4724, in which
NS have very long spin periods, are quasi-spherically accreting
SyXBs. In this case an ultra-strong magnetic field is not necessary
condition to explain the long spin periods.  GX 1+4 also probably is
a quasi-spherical wind-accreting source. However, we cannot identify
whether there are wind-fed accretion disks in 4U~1700+24, Sct~X-1,
IRXS~J180431.1-273932 and 2XMM J174016.0-290337.

\section*{Acknowledgements}
We acknowledge the referee for  constructive suggestions. LGL would
like to thank Christopher Tout at the Institute of Astronomy of
Cambridge University for his hospitality. This work was supported by
the National Natural Science Foundations of China (NSFC) grant
11063002 and 11163005, the Knowledge Innovation Program of the
Chinese Academy of Sciences(Grant No. KJCX2-YW-T09), National Basic
Research Program of China (973 Program 2009CB824800), Natural
Science Foundation of Xinjiang grant 2009211B01 and 2010211B05, and
Doctor Foundation of Xinjiang University (BS100106). KAP is
supported by RFBR grant 10-02-00951. LRY is supported by RFBR grant
10-02-00231 and the Program of the Praesidium of Russian academy of
sciences ``Nonstationary processes in the Universe objects''.

%%%%%%%%%%%%%%%%%%%%%%%%%%%%%%%%%%%%%%%%%%%%%%%%%%%%%%%%%%%
%%%%%%%%%%%%%%%%%%%%%%%%%%%%%%%%%%%%%%%%%%%%%%%%%%%%%%%%%%%%
%\bibliography{guoliangads}
%\begin{thebibliography}{99}
\bibliographystyle{mn2e}
\bibliography{lglmn1_pk,symbx_add}

\begin{thebibliography}{}

\bibitem[\protect\citeauthoryear{{Beskin}, {Gurevich} \& {Istomin}}{{Beskin}
  et~al.}{1993}]{Beskin1993}
{Beskin} V.~S.,  {Gurevich} A.~V.,    {Istomin} Y.~N.,  1993, {Physics of the
  pulsar magnetosphere}

\bibitem[\protect\citeauthoryear{{Bildsten}, {Chakrabarty}, {Chiu}, {Finger},
  {Koh}, {Nelson}, {Prince}, {Rubin}, {Scott}, {Stollberg}, {Vaughan}, {Wilson}
  \& {Wilson}}{{Bildsten} et~al.}{1997}]{Bildsten1997}
{Bildsten} L.,  {Chakrabarty} D.,  {Chiu} J.,  {Finger} M.~H.,  {Koh} D.~T.,
  {Nelson} R.~W.,  {Prince} T.~A.,  {Rubin} B.~C.,  {Scott} D.~M.,  {Stollberg}
  M.,  {Vaughan} B.~A.,  {Wilson} C.~A.,    {Wilson} R.~B.,  1997, ApJS, 113,
  367

\bibitem[\protect\citeauthoryear{{Bisnovatyi-Kogan} \&
  {Komberg}}{{Bisnovatyi-Kogan} \& {Komberg}}{1974}]{Bisnovatyi-Kogan1974}
{Bisnovatyi-Kogan} G.~S.,  {Komberg} B.~V.,  1974, Astron. Zh., 51, 373

\bibitem[\protect\citeauthoryear{{Bodaghee}, {Walter}, {Zurita Heras}, {Bird},
  {Courvoisier}, {Malizia}, {Terrier} \& {Ubertini}}{{Bodaghee}
  et~al.}{2006}]{Bodaghee2006}
{Bodaghee} A.,  {Walter} R.,  {Zurita Heras} J.~A.,  {Bird} A.~J.,
  {Courvoisier} T.,  {Malizia} A.,  {Terrier} R.,    {Ubertini} P.,  2006,
  A\&A, 447, 1027

\bibitem[\protect\citeauthoryear{{Bondi} \& {Hoyle}}{{Bondi} \&
  {Hoyle}}{1944}]{Bondi1944}
{Bondi} H.,  {Hoyle} F.,  1944, MNRAS, 104, 273

\bibitem[\protect\citeauthoryear{{Brown} \& {Lee}}{{Brown} \&
  {Lee}}{2004}]{Brown2004}
{Brown} G.~E.,  {Lee} C.,  2004, NewA, 9, 225

\bibitem[\protect\citeauthoryear{{B{\"u}ning} \& {Ritter}}{{B{\"u}ning} \&
  {Ritter}}{2004}]{Buning2004}
{B{\"u}ning} A.,  {Ritter} H.,  2004, A\&A, 423, 281

\bibitem[\protect\citeauthoryear{{Chakrabarty} \& {Roche}}{{Chakrabarty} \&
  {Roche}}{1997}]{Chakrabarty1997}
{Chakrabarty} D.,  {Roche} P.,  1997, ApJ, 489, 254

\bibitem[\protect\citeauthoryear{{Corbet}, {Sokoloski}, {Mukai}, {Markwardt} \&
  {Tueller}}{{Corbet} et~al.}{2008}]{Corbet2008}
{Corbet} R.~H.~D.,  {Sokoloski} J.~L.,  {Mukai} K.,  {Markwardt} C.~B.,
  {Tueller} J.,  2008, ApJ, 675, 1424

\bibitem[\protect\citeauthoryear{{Cordes}, {Kramer}, {Lazio}, {Stappers},
  {Backer} \& {Johnston}}{{Cordes} et~al.}{2004}]{Cordes2004}
{Cordes} J.~M.,  {Kramer} M.,  {Lazio} T.~J.~W.,  {Stappers} B.~W.,  {Backer}
  D.~C.,    {Johnston} S.,  2004, New Astronomy Review, 48, 1413

\bibitem[\protect\citeauthoryear{de Kool}{de~Kool}{1990}]{dek90}
de Kool M.,  1990, {\apj}, 358, 189

\bibitem[\protect\citeauthoryear{{Farrell}, {Gosling}, {Webb}, {Barret},
  {Rosen}, {Sakano} \& {Pancrazi}}{{Farrell} et~al.}{2010}]{Farrell2010}
{Farrell} S.~A.,  {Gosling} A.~J.,  {Webb} N.~A.,  {Barret} D.,  {Rosen} S.~R.,
   {Sakano} M.,    {Pancrazi} B.,  2010, A\&A, 523, A50+

\bibitem[\protect\citeauthoryear{{Faucher-Gigu{\`e}re} \&
  {Kaspi}}{{Faucher-Gigu{\`e}re} \& {Kaspi}}{2006}]{FGK2006}
{Faucher-Gigu{\`e}re} C.-A.,  {Kaspi} V.~M.,  2006, \apj, 643, 332

\bibitem[\protect\citeauthoryear{{Ghosh} \& {Lamb}}{{Ghosh} \&
  {Lamb}}{1978}]{Ghosh1978}
{Ghosh} P.,  {Lamb} F.~K.,  1978, ApJL, 223, L83

\bibitem[\protect\citeauthoryear{{Ghosh} \& {Lamb}}{{Ghosh} \&
  {Lamb}}{1979a}]{Ghosh1979a}
{Ghosh} P.,  {Lamb} F.~K.,  1979a, ApJ, 232, 259

\bibitem[\protect\citeauthoryear{{Ghosh} \& {Lamb}}{{Ghosh} \&
  {Lamb}}{1979b}]{Ghosh1979b}
{Ghosh} P.,  {Lamb} F.~K.,  1979b, ApJ, 234, 296

\bibitem[\protect\citeauthoryear{{Goldberg} \& {Mazeh}}{{Goldberg} \&
  {Mazeh}}{1994}]{1994A&A...282..801G}
{Goldberg} D.,  {Mazeh} T.,  1994, \aap, 282, 801

\bibitem[\protect\citeauthoryear{{Goldreich} \& {Reisenegger}}{{Goldreich} \&
  {Reisenegger}}{1992}]{Goldreich1992}
{Goldreich} P.,  {Reisenegger} A.,  1992, ApJ, 395, 250

\bibitem[\protect\citeauthoryear{{Gonz{\'a}lez-Gal{\'a}n}, {Kuulkers},
  {Kretschmar}, {Larsson}, {Postnov}, {Kochetkova} \&
  {Finger}}{{Gonz{\'a}lez-Gal{\'a}n} et~al.}{2012}]{GonzalezGalan2011}
{Gonz{\'a}lez-Gal{\'a}n} A.,  {Kuulkers} E.,  {Kretschmar} P.,  {Larsson} S.,
  {Postnov} K.,  {Kochetkova} A.,    {Finger} M.~H.,  2012, \aap, 537, A66

\bibitem[\protect\citeauthoryear{{Han}, {Podsiadlowski} \& {Eggleton}}{{Han}
  et~al.}{1995}]{Han1995b}
{Han} Z.,  {Podsiadlowski} P.,    {Eggleton} P.~P.,  1995, MNRAS, 272, 800

\bibitem[\protect\citeauthoryear{{Hansen} \& {Phinney}}{{Hansen} \&
  {Phinney}}{1997}]{Hansen1997}
{Hansen} B.~M.~S.,  {Phinney} E.~S.,  1997, MNRAS, 291, 569

\bibitem[\protect\citeauthoryear{{Harper}}{{Harper}}{1996}]{Harper1996}
{Harper} G.,  1996, in {R.~Pallavicini \& A.~K.~Dupree} ed., Cool Stars,
  Stellar Systems, and the Sun Vol.~109 of Astronomical Society of the Pacific
  Conference Series, {Mass loss and winds from cool giants}.
p.~481

\bibitem[\protect\citeauthoryear{{Hartman}, {Bhattacharya}, {Wijers} \&
  {Verbunt}}{{Hartman} et~al.}{1997}]{Hartman1997}
{Hartman} J.~W.,  {Bhattacharya} D.,  {Wijers} R.,    {Verbunt} F.,  1997,
  A\&A, 322, 477

\bibitem[\protect\citeauthoryear{{Hinkle}, {Fekel}, {Joyce}, {Wood}, {Smith} \&
  {Lebzelter}}{{Hinkle} et~al.}{2006}]{Hinkle2006}
{Hinkle} K.~H.,  {Fekel} F.~C.,  {Joyce} R.~R.,  {Wood} P.~R.,  {Smith} V.~V.,
    {Lebzelter} T.,  2006, ApJ, 641, 479

\bibitem[\protect\citeauthoryear{{Ho}, {Taam}, {Fryxell}, {Matsuda} \&
  {Koide}}{{Ho} et~al.}{1989}]{Ho_ea1989}
{Ho} C.,  {Taam} R.~E.,  {Fryxell} B.~A.,  {Matsuda} T.,    {Koide} H.,  1989,
  MNRAS, 238, 1447

\bibitem[\protect\citeauthoryear{{Hurley}, {Pols} \& {Tout}}{{Hurley}
  et~al.}{2000}]{Hurley2000}
{Hurley} J.~R.,  {Pols} O.~R.,    {Tout} C.~A.,  2000, MNRAS, 315, 543

\bibitem[\protect\citeauthoryear{{Hurley}, {Tout} \& {Pols}}{{Hurley}
  et~al.}{2002}]{Hurley2002}
{Hurley} J.~R.,  {Tout} C.~A.,    {Pols} O.~R.,  2002, MNRAS, 329, 897

\bibitem[\protect\citeauthoryear{{Iben} Jr. \& {Tutukov}}{{Iben} \&
  {Tutukov}}{1996}]{1996ApJS..105..145I}
{Iben} Jr. I.,  {Tutukov} A.~V.,  1996, \apjs, 105, 145

\bibitem[\protect\citeauthoryear{{Iben} Jr., {Tutukov} \& {Yungelson}}{{Iben}
  et~al.}{1995}]{1995ApJS..100..233I}
{Iben} Jr. I.,  {Tutukov} A.~V.,    {Yungelson} L.~R.,  1995, \apjs, 100, 233

\bibitem[\protect\citeauthoryear{{Illarionov} \& {Sunyaev}}{{Illarionov} \&
  {Sunyaev}}{1975}]{Illarionov1975}
{Illarionov} A.~F.,  {Sunyaev} R.~A.,  1975, A\&A, 39, 185

\bibitem[\protect\citeauthoryear{{Ivanova}, {Heinke}, {Rasio}, {Belczynski} \&
  {Fregeau}}{{Ivanova} et~al.}{2008}]{Ivanova2008}
{Ivanova} N.,  {Heinke} C.~O.,  {Rasio} F.~A.,  {Belczynski} K.,    {Fregeau}
  J.~M.,  2008, MNRAS, 386, 553

\bibitem[\protect\citeauthoryear{{Kaplan}, {Levine}, {Chakrabarty}, {Morgan},
  {Erb}, {Gaensler}, {Moon} \& {Cameron}}{{Kaplan} et~al.}{2007}]{Kaplan2007}
{Kaplan} D.~L.,  {Levine} A.~M.,  {Chakrabarty} D.,  {Morgan} E.~H.,  {Erb}
  D.~K.,  {Gaensler} B.~M.,  {Moon} D.,    {Cameron} P.~B.,  2007, ApJ, 661,
  437

\bibitem[\protect\citeauthoryear{{Katz}}{{Katz}}{1975}]{Katz1975}
{Katz} J.~I.,  1975, Nat, 253, 698

\bibitem[\protect\citeauthoryear{{Kiel} \& {Hurley}}{{Kiel} \&
  {Hurley}}{2006}]{Kiel2006}
{Kiel} P.~D.,  {Hurley} J.~R.,  2006, MNRAS, 369, 1152

\bibitem[\protect\citeauthoryear{{Kiel}, {Hurley}, {Bailes} \& {Murray}}{{Kiel}
  et~al.}{2008}]{Kiel2008}
{Kiel} P.~D.,  {Hurley} J.~R.,  {Bailes} M.,    {Murray} J.~R.,  2008, MNRAS,
  388, 393

\bibitem[\protect\citeauthoryear{{Kraicheva}, {Popova}, {Tutukov} \&
  {Yungelson}}{{Kraicheva} et~al.}{1989}]{1989Ap.....30..323K}
{Kraicheva} Z.~T.,  {Popova} E.~I.,  {Tutukov} A.~V.,    {Yungelson} L.~R.,
  1989, Astrophysics, 30, 323

\bibitem[\protect\citeauthoryear{{Lipunov}}{{Lipunov}}{1982}]{Lipunov1982}
{Lipunov} V.~M.,  1982, Soviet Astronomy, 26, 54

\bibitem[\protect\citeauthoryear{{Lipunov}}{{Lipunov}}{1987}]{Lipunov1987}
{Lipunov} V.~M.,  1987, Ap\&SS, 132, 1

\bibitem[\protect\citeauthoryear{{Lipunov}, {B{\"o}rner} \& {Wadhwa}}{{Lipunov}
  et~al.}{1992}]{Lipunov1992}
{Lipunov} V.~M.,  {B{\"o}rner} G.,    {Wadhwa} R.~S.,  1992, {Astrophysics of
  Neutron Stars}

\bibitem[\protect\citeauthoryear{{Lipunov} \& {Postnov}}{{Lipunov} \&
  {Postnov}}{1988}]{Lipunov1988}
{Lipunov} V.~M.,  {Postnov} K.~A.,  1988, Ap\&SS, 145, 1

\bibitem[\protect\citeauthoryear{{Liu}, {van Paradijs} \& {van den
  Heuvel}}{{Liu} et~al.}{2006}]{Liu2006}
{Liu} Q.~Z.,  {van Paradijs} J.,    {van den Heuvel} E.~P.~J.,  2006, A\&A,
  455, 1165

\bibitem[\protect\citeauthoryear{{Liu}, {van Paradijs} \& {van den
  Heuvel}}{{Liu} et~al.}{2007}]{Liu2007}
{Liu} Q.~Z.,  {van Paradijs} J.,    {van den Heuvel} E.~P.~J.,  2007, A\&A,
  469, 807

\bibitem[\protect\citeauthoryear{{Lovelace}, {Romanova} \&
  {Bisnovatyi-Kogan}}{{Lovelace} et~al.}{1995}]{Lovelace1995}
{Lovelace} R.~V.~E.,  {Romanova} M.~M.,    {Bisnovatyi-Kogan} G.~S.,  1995,
  MNRAS, 275, 244

\bibitem[\protect\citeauthoryear{{Lovelace}, {Romanova} \&
  {Bisnovatyi-Kogan}}{{Lovelace} et~al.}{1999}]{Lovelace1999}
{Lovelace} R.~V.~E.,  {Romanova} M.~M.,    {Bisnovatyi-Kogan} G.~S.,  1999,
  ApJ, 514, 368

\bibitem[\protect\citeauthoryear{{L{\"u}}, {Yungelson} \& {Han}}{{L{\"u}}
  et~al.}{2006}]{Lu2006}
{L{\"u}} G.,  {Yungelson} L.,    {Han} Z.,  2006, MNRAS, 372, 1389

\bibitem[\protect\citeauthoryear{{L{\"u}}, {Zhu}, {Han} \& {Wang}}{{L{\"u}}
  et~al.}{2008}]{Lu2008}
{L{\"u}} G.,  {Zhu} C.,  {Han} Z.,    {Wang} Z.,  2008, ApJ, 683, 990

\bibitem[\protect\citeauthoryear{{L{\"u}}, {Zhu}, {Wang}, {Huo} \&
  {Yang}}{{L{\"u}} et~al.}{2011}]{Lu2011}
{L{\"u}} G.,  {Zhu} C.,  {Wang} Z.,  {Huo} W.,    {Yang} Y.,  2011, MNRAS, 413,
  L11

\bibitem[\protect\citeauthoryear{{L{\"u}}, {Zhu}, {Wang} \& {Wang}}{{L{\"u}}
  et~al.}{2009}]{Lu2009}
{L{\"u}} G.,  {Zhu} C.,  {Wang} Z.,    {Wang} N.,  2009, MNRAS, 396, 1086

\bibitem[\protect\citeauthoryear{{Lutovinov}, {Revnivtsev}, {Gilfanov},
  {Shtykovskiy}, {Molkov} \& {Sunyaev}}{{Lutovinov}
  et~al.}{2005}]{Lutovinov2005}
{Lutovinov} A.,  {Revnivtsev} M.,  {Gilfanov} M.,  {Shtykovskiy} P.,  {Molkov}
  S.,    {Sunyaev} R.,  2005, A\&A, 444, 821

\bibitem[\protect\citeauthoryear{{Lutovinov}, {Tsygankov} \&
  {Chernyakova}}{{Lutovinov} et~al.}{2012}]{Lutovinov_ea2012}
{Lutovinov} A.,  {Tsygankov} S.,    {Chernyakova} M.,  2012, ArXiv e-prints

\bibitem[\protect\citeauthoryear{{Marcu}, {F{\"u}rst}, {Pottschmidt},
  {Grinberg}, {M{\"u}ller}, {Wilms}, {Postnov}, {Corbet}, {Markwardt} \&
  {Cadolle Bel}}{{Marcu} et~al.}{2011}]{Marcu2011}
{Marcu} D.~M.,  {F{\"u}rst} F.,  {Pottschmidt} K.,  {Grinberg} V.,
  {M{\"u}ller} S.,  {Wilms} J.,  {Postnov} K.~A.,  {Corbet} R.~H.~D.,
  {Markwardt} C.~B.,    {Cadolle Bel} M.,  2011, ApJL, 742, L11

\bibitem[\protect\citeauthoryear{{Masetti}, {Dal Fiume}, {Cusumano}, {Amati},
  {Bartolini}, {Del Sordo}, {Frontera}, {Guarnieri}, {Orlandini}, {Palazzi},
  {Parmar}, {Piccioni} \& {Santangelo}}{{Masetti} et~al.}{2002}]{Masetti2002}
{Masetti} N.,  {Dal Fiume} D.,  {Cusumano} G.,  {Amati} L.,  {Bartolini} C.,
  {Del Sordo} S.,  {Frontera} F.,  {Guarnieri} A.,  {Orlandini} M.,  {Palazzi}
  E.,  {Parmar} A.~N.,  {Piccioni} A.,    {Santangelo} A.,  2002, A\&A, 382,
  104

\bibitem[\protect\citeauthoryear{{Masetti}, {Landi}, {Pretorius}, {Sguera},
  {Bird}, {Perri}, {Charles}, {Kennea}, {Malizia} \& {Ubertini}}{{Masetti}
  et~al.}{2007}]{Masetti2007}
{Masetti} N.,  {Landi} R.,  {Pretorius} M.~L.,  {Sguera} V.,  {Bird} A.~J.,
  {Perri} M.,  {Charles} P.~A.,  {Kennea} J.~A.,  {Malizia} A.,    {Ubertini}
  P.,  2007, A\&A, 470, 331

\bibitem[\protect\citeauthoryear{{Masetti}, {Munari}, {Henden}, {Page},
  {Osborne} \& {Starrfield}}{{Masetti} et~al.}{2011}]{Masetti2011}
{Masetti} N.,  {Munari} U.,  {Henden} A.~A.,  {Page} K.~L.,  {Osborne} J.~P.,
   {Starrfield} S.,  2011, ArXiv e-prints

\bibitem[\protect\citeauthoryear{{Masetti}, {Orlandini}, {Palazzi}, {Amati} \&
  {Frontera}}{{Masetti} et~al.}{2006}]{Masetti2006}
{Masetti} N.,  {Orlandini} M.,  {Palazzi} E.,  {Amati} L.,    {Frontera} F.,
  2006, A\&A, 453, 295

\bibitem[\protect\citeauthoryear{{Miller} \& {Scalo}}{{Miller} \&
  {Scalo}}{1979}]{1979ApJS...41..513M}
{Miller} G.~E.,  {Scalo} J.~M.,  1979, \apjs, 41, 513

\bibitem[\protect\citeauthoryear{{Miyaji}, {Nomoto}, {Yokoi} \&
  {Sugimoto}}{{Miyaji} et~al.}{1980}]{Miyaji1980}
{Miyaji} S.,  {Nomoto} K.,  {Yokoi} K.,    {Sugimoto} D.,  1980, PASJ, 32, 303

\bibitem[\protect\citeauthoryear{{M\"{u}rset}, {Wolff} \&
  {Jordan}}{{M\"{u}rset} et~al.}{1997}]{Murset1997}
{M\"{u}rset} U.,  {Wolff} B.,    {Jordan} S.,  1997, A\&A, 319, 201

\bibitem[\protect\citeauthoryear{{Nespoli}, {Fabregat} \&
  {Mennickent}}{{Nespoli} et~al.}{2010}]{Nespoli2010}
{Nespoli} E.,  {Fabregat} J.,    {Mennickent} R.~E.,  2010, A\&A, 516, A94+

\bibitem[\protect\citeauthoryear{{Nucita}, {Carpano} \& {Guainazzi}}{{Nucita}
  et~al.}{2007}]{Nucita2007}
{Nucita} A.~A.,  {Carpano} S.,    {Guainazzi} M.,  2007, A\&A, 474, L1

\bibitem[\protect\citeauthoryear{{Os{\l}owski}, {Bulik}, {Gondek-Rosi{\'n}ska}
  \& {Belczy{\'n}ski}}{{Os{\l}owski} et~al.}{2011}]{Oslowski2011}
{Os{\l}owski} S.,  {Bulik} T.,  {Gondek-Rosi{\'n}ska} D.,    {Belczy{\'n}ski}
  K.,  2011, MNRAS, pp 103--+

\bibitem[\protect\citeauthoryear{{Patel}, {Kouveliotou}, {Tennant}, {Woods},
  {King}, {Finger}, {Ubertini}, {Winkler}, {Courvoisier}, {van der Klis},
  {Wachter}, {Gaensler} \& {Phillips}}{{Patel} et~al.}{2004}]{Patel2004}
{Patel} S.~K.,  {Kouveliotou} C.,  {Tennant} A.,  {Woods} P.~M.,  {King} A.,
  {Finger} M.~H.,  {Ubertini} P.,  {Winkler} C.,  {Courvoisier} T.,  {van der
  Klis} M.,  {Wachter} S.,  {Gaensler} B.~M.,    {Phillips} C.~J.,  2004, ApJL,
  602, L45

\bibitem[\protect\citeauthoryear{{Patel}, {Zurita}, {Del Santo}, {Finger},
  {Kouveliotou}, {Eichler}, {G{\"o}{\u g}{\"u}{\c s}}, {Ubertini}, {Walter},
  {Woods}, {Wilson}, {Wachter} \& {Bazzano}}{{Patel} et~al.}{2007}]{Patel2007}
{Patel} S.~K.,  {Zurita} J.,  {Del Santo} M.,  {Finger} M.,  {Kouveliotou} C.,
  {Eichler} D.,  {G{\"o}{\u g}{\"u}{\c s}} E.,  {Ubertini} P.,  {Walter} R.,
  {Woods} P.,  {Wilson} C.~A.,  {Wachter} S.,    {Bazzano} A.,  2007, ApJ, 657,
  994

\bibitem[\protect\citeauthoryear{{Pfahl}, {Rappaport} \&
  {Podsiadlowski}}{{Pfahl} et~al.}{2002}]{Pfahl2002}
{Pfahl} E.,  {Rappaport} S.,    {Podsiadlowski} P.,  2002, ApJ, 573, 283

\bibitem[\protect\citeauthoryear{{Pfahl}, {Rappaport} \&
  {Podsiadlowski}}{{Pfahl} et~al.}{2003}]{Pfahl2003}
{Pfahl} E.,  {Rappaport} S.,    {Podsiadlowski} P.,  2003, ApJ, 597, 1036

\bibitem[\protect\citeauthoryear{{Podsiadlowski}, {Langer}, {Poelarends},
  {Rappaport}, {Heger} \& {Pfahl}}{{Podsiadlowski}
  et~al.}{2004}]{Podsiadlowski2004}
{Podsiadlowski} P.,  {Langer} N.,  {Poelarends} A.~J.~T.,  {Rappaport} S.,
  {Heger} A.,    {Pfahl} E.,  2004, ApJ, 612, 1044

\bibitem[\protect\citeauthoryear{{Popov}, {Pons}, {Miralles}, {Boldin} \&
  {Posselt}}{{Popov} et~al.}{2010}]{Popov_ea2010}
{Popov} S.~B.,  {Pons} J.~A.,  {Miralles} J.~A.,  {Boldin} P.~A.,    {Posselt}
  B.,  2010, \mnras, 401, 2675

\bibitem[\protect\citeauthoryear{{Postnov}, {Shakura}, {Kochetkova} \&
  {Hjalmarsdotter}}{{Postnov} et~al.}{2011}]{PKchia}
{Postnov} K.,  {Shakura} N.,  {Kochetkova} A.,    {Hjalmarsdotter} L.,  2011,
  ArXiv e-prints

\bibitem[\protect\citeauthoryear{{Pringle} \& {Rees}}{{Pringle} \&
  {Rees}}{1972}]{Pringle1972}
{Pringle} J.~E.,  {Rees} M.~J.,  1972, A\&A, 21, 1

\bibitem[\protect\citeauthoryear{{Rappaport}, {Verbunt} \& {Joss}}{{Rappaport}
  et~al.}{1983}]{1983ApJ...275..713R}
{Rappaport} S.,  {Verbunt} F.,    {Joss} P.~C.,  1983, \apj, 275, 713

\bibitem[\protect\citeauthoryear{{Revnivtsev}, {Postnov}, {Kuranov} \&
  {Ritter}}{{Revnivtsev} et~al.}{2011}]{Revnivtsev2011}
{Revnivtsev} M.,  {Postnov} K.,  {Kuranov} A.,    {Ritter} H.,  2011, A\&A,
  526, A94+

\bibitem[\protect\citeauthoryear{{Romanova}, {Ustyugova}, {Koldoba} \&
  {Lovelace}}{{Romanova} et~al.}{2002}]{Romanova2002}
{Romanova} M.~M.,  {Ustyugova} G.~V.,  {Koldoba} A.~V.,    {Lovelace} R.~V.~E.,
   2002, ApJ, 578, 420

\bibitem[\protect\citeauthoryear{{Romanova}, {Ustyugova}, {Koldoba} \&
  {Lovelace}}{{Romanova} et~al.}{2004}]{Romanova2004}
{Romanova} M.~M.,  {Ustyugova} G.~V.,  {Koldoba} A.~V.,    {Lovelace} R.~V.~E.,
   2004, ApJL, 616, L151

\bibitem[\protect\citeauthoryear{{Romanova}, {Ustyugova}, {Koldoba} \&
  {Lovelace}}{{Romanova} et~al.}{2005}]{Romanova2005}
{Romanova} M.~M.,  {Ustyugova} G.~V.,  {Koldoba} A.~V.,    {Lovelace} R.~V.~E.,
   2005, ApJL, 635, L165

\bibitem[\protect\citeauthoryear{{Romanova}, {Ustyugova}, {Koldoba}, {Wick} \&
  {Lovelace}}{{Romanova} et~al.}{2003}]{Romanova2003}
{Romanova} M.~M.,  {Ustyugova} G.~V.,  {Koldoba} A.~V.,  {Wick} J.~V.,
  {Lovelace} R.~V.~E.,  2003, ApJ, 595, 1009

\bibitem[\protect\citeauthoryear{{Ruderman}}{{Ruderman}}{1991}]{Ruderman1991}
{Ruderman} M.,  1991, ApJ, 366, 261

\bibitem[\protect\citeauthoryear{{Shakura}, {Postnov}, {Kochetkova} \&
  {Hjalmarsdotter}}{{Shakura} et~al.}{2012}]{Shakura2011}
{Shakura} N.,  {Postnov} K.,  {Kochetkova} A.,    {Hjalmarsdotter} L.,  2012,
  \mnras, 420, 216

\bibitem[\protect\citeauthoryear{{Shakura} \& {Sunyaev}}{{Shakura} \&
  {Sunyaev}}{1973}]{Shakura1973}
{Shakura} N.~I.,  {Sunyaev} R.~A.,  1973, A\&A, 24, 337

\bibitem[\protect\citeauthoryear{{Thompson}, {Tomsick}, {Rothschild}, {in't
  Zand} \& {Walter}}{{Thompson} et~al.}{2006}]{Thompson2006}
{Thompson} T.~W.~J.,  {Tomsick} J.~A.,  {Rothschild} R.~E.,  {in't Zand}
  J.~J.~M.,    {Walter} R.,  2006, ApJ, 649, 373

\bibitem[\protect\citeauthoryear{{Tutukov} \& {Yungelson}}{{Tutukov} \&
  {Yungelson}}{1976}]{ty76}
{Tutukov} A.~V.,  {Yungelson} L.~R.,  1976, \astroph, 12, 521

\bibitem[\protect\citeauthoryear{{van den Heuvel} \& {Bitzaraki}}{{van den
  Heuvel} \& {Bitzaraki}}{1995}]{Heuvel1995}
{van den Heuvel} E.~P.~J.,  {Bitzaraki} O.,  1995, A\&A, 297, L41+

\bibitem[\protect\citeauthoryear{{Verbunt} \& {Zwaan}}{{Verbunt} \&
  {Zwaan}}{1981}]{Verbunt1981}
{Verbunt} F.,  {Zwaan} C.,  1981, A\&A, 100, L7

\bibitem[\protect\citeauthoryear{{Webbink}}{{Webbink}}{1984}]{Webbink1984}
{Webbink} R.~F.,  1984, ApJ, 277, 355

\bibitem[\protect\citeauthoryear{{Webbink}}{{Webbink}}{1988}]{Webbink1988}
{Webbink} R.~F.,  1988, in {Mikolajewska, J., Friedjung, M., Kenyon, S.~J., \&
  Viotti, R. } ed., The Symbiotic Phenomenon, Proceedings of IAU Colloq.~103,
  held in Torun, Poland, 18-20 August, 1987 (ASSL 145). {The Formation and
  Evolution of Symbiotic Stars}.
Kluwer Academic Publishers, Dordrecht, p.~311

\bibitem[\protect\citeauthoryear{{Yungelson}, {Livio}, {Tutukov} \&
  {Kenyon}}{{Yungelson} et~al.}{1995}]{Yungelson1995}
{Yungelson} L.,  {Livio} M.,  {Tutukov} A.,    {Kenyon} S.~J.,  1995, ApJ, 447,
  656

\bibitem[\protect\citeauthoryear{{Yungelson}, {Tutukov} \& {Livio}}{{Yungelson}
  et~al.}{1993}]{Yungelson1993}
{Yungelson} L.~R.,  {Tutukov} A.~V.,    {Livio} M.,  1993, ApJ, 418, 794

\bibitem[\protect\citeauthoryear{{Zhu}, {L{\"u}}, {Wang} \& {Wang}}{{Zhu}
  et~al.}{2012}]{Zhu2012}
{Zhu} C.,  {L{\"u}} G.,  {Wang} Z.,    {Wang} N.,  2012, \pasp, 124, 195

\end{thebibliography}
%\end{thebibliography}

\bsp

\label{lastpage}

\end{document}